\documentclass[prd,aps,floats,floatfix,superscriptaddress,preprintnumbers,
showpacs,eqsecnum,nofootinbib,twocolumn]{revtex4}
\usepackage{latexsym,array,theorem,mathrsfs,bm,float}

\usepackage{psfrag}
\usepackage{amsfonts,amsmath,amssymb,latexsym,array,afterpage,theorem,mathrsfs,bm,float,epsfig,color,graphicx,tabularx,here,multirow}
\newcommand{\nn}{\nonumber \\}
\newcommand{\bea}{\begin{eqnarray}}
\newcommand{\ena}{\end{eqnarray}}
\newcommand{\be}{\begin{equation}}
\newcommand{\ee}{\end{equation}}
\newcommand{\beann}{\begin{eqnarray*}}
\newcommand{\enann}{\end{eqnarray*}}

\newcommand{\ma}[1]{\mbox{$\mathcal{#1}$}}

\newcommand{\calhR}[1]{\raisebox{2ex}{\tiny ({\em h})}\hspace{-0.8em}{\ma R}}

\begin{document}

%<<<<<<<<<<<<< TITLE >>>>>>>>>>>>>>>%
\title{
Resolving Hubble Tension with Quintom Dark Energy Model
}

%<<<<<<<<<<<<< AUTHOR >>>>>>>>>>>>>>>%

\author{Sirachak {Panpanich}}
\email{sirachakp-at-gmail.com}
\address{High Energy Physics Theory Group, Department of Physics, 
Faculty of Science, Chulalongkorn University, Phayathai Rd., 
Bangkok 10330, Thailand}
\author{Piyabut {Burikham}}
\email{piyabut-at-gmail.com}
\address{High Energy Physics Theory Group, Department of Physics, 
Faculty of Science, Chulalongkorn University, Phayathai Rd., 
Bangkok 10330, Thailand}
\author{Supakchai {Ponglertsakul}}
\email{supakchai.p-at-gmail.com}
\address{Department of Physics and Astronomy, Sejong University, Seoul 05006, Republic of Korea}
\author{Lunchakorn {Tannukij}}
\email{l_tannukij-at-hotmail.com}
\address{Theoretical and Computational Physics Group,Theoretical and Computational Science Center(TaCS), Faculty of Science,King Mongkut's University of Technology Thonburi, Prachautid Rd., Bangkok 10140, Thailand}

%<<<<<<<<<<<<< ADDRESS >>>>>>>>>>>>>>>%

%<<<<<<<<<<<<< DATE >>>>>>>>>>>>>>>%
\date{\today}

%======================================%
%<<<<<<<<<<<<< ABSTRACT >>>>>>>>>>>>>>>%
%======================================%
%%%%%%%%%%%%%%%%%%%%%%%%%%%%%%%%%%%%%%%%%%%
%%%%%%%%%%%%%%%%%%%%%%%%%%%%%%%%%%%%%%%%%%%
%%%%%%%%%%%%%%%%%%%%%%%%%%%%%%%%%%%%%%%%%%%
%%%%%%%%%%%%%%%%%%%%%%%%%%%%%%%%%%%%%%%%%%%
\begin{abstract}

Recent low-redshift observations give value of the present-time Hubble parameter $H_{0}\simeq 74~\rm{km s}^{-1}\rm{Mpc}^{-1}$, roughly 10\% higher than the predicted value $H_{0}=67.4~\rm{km s}^{-1}\rm{Mpc}^{-1}$ from Planck's observations of the Cosmic Microwave Background radiation~(CMB) and the $\Lambda$CDM model. Phenomenologically, we show that by adding an extra component X with negative density in the Friedmann equation, it can relieve the Hubble tension without changing the Planck's constraint on the matter and dark energy densities. For the extra negative density to be sufficiently small, its equation-of-state parameter must satisfy $1/3\leq w_{X}\leq1$. We propose a quintom model of two scalar fields that realizes this condition and potentially alleviate the Hubble tension.  One scalar field acts as a quintessence while another ``phantom'' scalar conformally couples to matter in such a way that viable cosmological scenario can be achieved.  The model depends only on two parameters, $\lambda_{\phi}$ and $\delta$ which represent rolling tendency of the self-interacting potential of the quintessence and the strength of conformal phantom-matter coupling respectively. The toy quintom model with $H_{0}=73.4~\rm{km s}^{-1}\rm{Mpc}^{-1}$~(Quintom I) gives good Supernovae-Ia luminosity fits, decent $r_{\rm BAO}$ fit, but slightly small acoustic multipole $\ell_{A}=285.54$. Full parameter scan reveals that quintom model provide better model than the $\Lambda$CDM model in certain region of the parameter space, $0.02<\delta<0.10, \Omega_{m}^{(0)}<0.31$, while significantly relieving Hubble tension even though not completely resolving it. A benchmark quintom model, Quintom II, is presented as an example.

\end{abstract}

\maketitle

%======================================%
%<<<<<<<<<<<< SECTION I  >>>>>>>>>>>>>>%
%======================================%
%%%%%%%%%%%%%%%%%%%%%%%%%%%%%%%%%%%%%%%%%%%
%%%%%%%%%%%%%%%%%%%%%%%%%%%%%%%%%%%%%%%%%%%
%%%%%%%%%%%%%%%%%%%%%%%%%%%%%%%%%%%%%%%%%%%
%%%%%%%%%%%%%%%%%%%%%%%%%%%%%%%%%%%%%%%%%%%
\section{Introduction}
%%%%%%%%%%%%%%%%%%%%%%%%%%%%%%%%%%%%%%%%%%%%%%%%%%%
%%%%%%%%%%%%%%%%%%%%%%%%%%%%%%%%%%%%%%%%%%%%%%%%%%%
%%%%%%%%%%%%%%%%%%%%%%%%%%%%%%%%%%%%%%%%%%%%%%%%%%%

After discovery of the accelerated expansion of the Universe \cite{Riess:1998cb,Perlmutter:1998np}, a number of hypotheses have been proposed to solve the dark energy problem, such as Horndeski theories \cite{Horndeski:1974wa,Deffayet:2011gz,Kobayashi:2011nu}, generalized proca theories \cite{Heisenberg:2014rta,DeFelice:2016yws}, or a ghost-free massive gravity \cite{deRham:2010kj,deRham:2010ik}. However, the discovery of gravitational waves GW170817~\cite{Abbott:2016blz} severely constrains these modified gravity models~\cite{TheLIGOScientific:2017qsa,Baker:2017hug,Sakstein:2017xjx}. The simplest standard model of cosmology, without introducing any new gravitational degrees of freedom, is the $\Lambda$CDM.  With ``minimal'' proposal of dark matter and dark energy components, it can explain the accelerated expansion of the Universe as well as other observational data reasonably well until recently~\cite{Ade:2015xua}.  A number of low-redshift observations reveals that there are discrepancies between the values of the Hubble parameter at the present time $H_{0}$ from observations of Cepheids in the Large Magellanic Cloud (LMC)~\cite{Riess:2019cxk}, the gravitational lensing of quasars measurement~\cite{Wong:2019kwg,Chen:2019ejq}, and the predicted value from the Planck CMB data within the $\Lambda$CDM~(Note that an intermediate value is originally found by using Red Giants as the distance ladder~\cite{Freedman:2019jwv} but is later corrected to the value consistent with the low-redshift measurements~\cite{Yuan:2019npk}). Since the difference between $H_0$ is roughly $5 \sigma$ in significance, this means that the standard model of cosmology, the $\Lambda$CDM, may not be correct. There exists tension between $H_{0}$ predicted from the early Universe and those directly measured at low redshifts.

Many ideas have been proposed to resolve the Hubble tension, such as the modified gravity and brane model~\cite{Renk:2017rzu,Nunes:2018xbm,Desmond:2019ygn,Alam:2016wpf}, the gravitational and vacuum~\cite{Khosravi:2017hfi,DiValentino:2017rcr,Banihashemi:2018has} phase transition, the early dark energy~\cite{Poulin:2018cxd}, the dark matter decay~\cite{Vattis:2019efj}, the dark sector interaction~\cite{Kumar:2019wfs,Yang:2018euj}, the neutrino self-interaction~\cite{Kreisch:2019yzn}, the phenomenological dark energy~\cite{Li:2019yem} and the negative cosmological constant~\cite{Dutta:2018vmq}~(see however \cite{Visinelli:2019qqu}). In this work, we demonstrate that the usual Friedmann equation allows a higher value of $H_{0}=74.03~\rm{km~s}^{-1}\rm{Mpc}^{-1}$ while keeping the matter contribution to $31\%$ and the dark energy contribution to $69\%$, provided that an extra component with very small negative density is introduced.  The negative component must be a very small fraction to the total density of the Universe otherwise it would have been detected~(see Ref.~\cite{Panpanich:2018cxo} however, for possible galactic effects of small negative density to the rotation curves).  As a theoretical model of such possibility, we propose a modified quintom model~\cite{Guo:2004fq} to realize a negative-density component required by the Friedmann equation phenomenologically. The quintom model consists of a quintessence scalar field and a phantom scalar field.  The model can provide dark energy with phantom divide crossing, while a late-time solution is still stable. Using two scalar fields for dark energy is not a new novel~(see e.g. \cite{Elizalde:2004mq,Elizalde:2008yf,Alexander:2019rsc}, it is also shown in Ref.~\cite{Colgain:2019joh} that a model with only one quintessence scalar and a cosmological constant makes Hubble tension worse), a model called a gravitational scalar-tensor theory also possesses two scalar fields \cite{Naruko:2015zze,Saridakis:2016ahq}. It is interesting to see whether the phantom scalar field of the quintom model matches with the required negative density and could alleviate the Hubble tension.

This work is organized as follows. Section \ref{GenPhen} generally discusses the physical requirement of the extra component $X$ to coexist within the standard Friedmann model in order to alleviate the Hubble tension.  In Section \ref{basiceq} we propose a modified quintom model with scalar-matter coupling that realizes the negative density requirement, giving the right $H_{0}$ while keeping the density parameters $\Omega_{m}^{(0)}\simeq 0.31, \Omega_{DE}^{(0)}\simeq 0.69$ consistent with Planck's early Universe constraints. We use a dynamical system approach to find cosmological solutions of the modified quintom model in Section \ref{dynamical}. Section~\ref{numer} contains the numerical analysis of the quintom model, yielding realistic cosmological solution. Section \ref{luminosity} compares theoretical prediction of our model with the observational data and Section \ref{conclusions} summarizes our work.

\section{General Phenomenology}  \label{GenPhen}

In this section, a general physical condition is discussed based on the Friedmann equation with one extra component in addition to the standard $\Lambda$CDM model. In this approach it is assumed that the Planck's constraints from the early Universe on $H_{0}$ is valid, i.e., $\Omega_{m}^{(0)}=0.31, \Omega_{\Lambda}^{(0)}=0.69$ and very small contributions from other components at the present day.  

\begin{figure}[h]
	\centerline{\includegraphics[width=7cm]{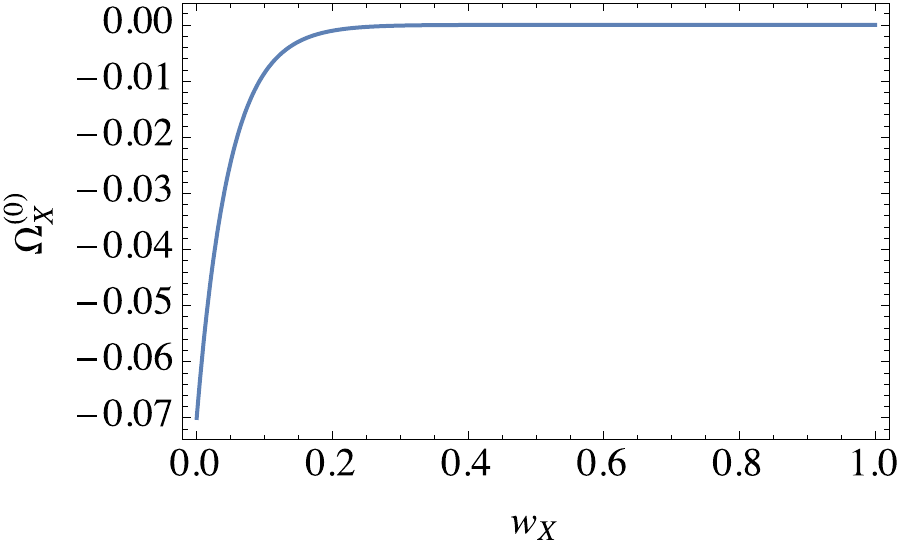}}
	\caption{A relation between $\Omega_{X}^{(0)}$ and $w_{X}$ from Eq. (\ref{pheneq}) for $\Omega_m^{(0)} = 0.308$, $\Omega_r^{(0)} = 9.2 \times 10^{-5}$, and $\Omega_{\Lambda =DE}^{(0)} =1-\Omega_{r}^{(0)}-\Omega_m^{(0)}-\Omega_X^{(0)}, w_{DE}=-1$.}
\label{omegaX}
\end{figure}

Using a density parameter, $\Omega_i \equiv \rho_i/\rho_c$, where $\rho_c\equiv 3H^{2}/8\pi G$ is the critical density of the Universe, the generalized Friedmann equation for the spatially flat Universe is given by
\bea  \label{pheneq}
H^2 (z) &=&  H^2_{0} \Big[\Omega_r^{(0)} (1+z)^4 + \Omega_m^{(0)} (1+z)^3 \nonumber\\
&&+ \Omega_{DE}^{(0)} \exp\left(3\int_0^z \frac{1+w_{DE}(z)}{1+z} dz\right) \nonumber \\
& & + \Omega_{X}^{(0)} \exp\left(3\int_0^z \frac{1+w_{X}(z)}{1+z} dz\right) \Big] \,,
\ena
where the subscript $r,m,DE,X$ represents radiation, matter, dark energy, and the extra unknown component $X$ respectively.
Notation ``$(0)$" denotes the present value at zero redshift. In the $\Lambda$CDM model with $w_{DE}=-1$, observational data from the CMB and high-redshifts prefers $\Omega_m^{(0)} = 0.308$, $\Omega_r^{(0)} = 9.2 \times 10^{-5}$, and $\Omega_{\Lambda =DE}^{(0)} = 0.692$, and $H_{0} = 67.4 ~ {\rm km ~s}^{-1}{\rm Mpc}^{-1}$ \cite{Ade:2015xua,Aghanim:2018eyx}. We can thus calculate $H(z)$ at the last-scattering surface ($z \approx 1100$) to be approximately $1.57537\times 10^{6}~\rm{km ~s}^{-1}{\rm Mpc}^{-1}$.  In order to address the Hubble tension where the value of $H$ at small $z$ is relatively large compared to the Planck value $H_{0}=67.4~\text{km s}^{-1}{\rm Mpc}^{-1}$, we use $H(z=1100)=1.57537\times 10^{6}~\rm{km ~s}^{-1}{\rm Mpc}^{-1}$ and $H_{0}=74.03~\rm{km ~s}^{-1}{\rm Mpc}^{-1}$ to find the physical constraint on the extra unknown component $X$ from Eq.~(\ref{pheneq}).  The allowed values of $w_{X},\Omega_{X}^{(0)}$ are shown in Fig.~\ref{omegaX}, assuming the simplest case where $w_{X}$ is constant.

Interestingly, {\it negative energy density} $\Omega_{X}^{(0)}<0$ is required. According to Fig. \ref{omegaX}, the negative density cannot be a negative mass ($w_{X} \approx 0$) otherwise $\Omega_{X}^{(0)}$ is too large $\Omega_{X}^{(0)}\simeq -0.07$, and it should have been observed. For $1/3\leq w_{X}\leq 1$, however, the amount of the extra component $X$ is very small, i.e., $-0.0000636042 \leq \Omega_X^{(0)} \leq -5.247 \times 10^{-11}$.  Thus in order to solve the Hubble tension problem we require a negative density with $1/3 \leq w_{X} \leq 1$ without modification on the CMB observational data.

This particular phenomenological model actually fails when extrapolated back to early times due to the overpopulation of the negative energy. At $z=1100$, the density parameter becomes $\Omega_{X}= -0.23$ which should be excluded by constraints on the power spectrum of the CMB on the early DE. We need a more realistic model with cosmological evolution that suppressed the early DE population, and satisfies the early-time constraints from the CMB and late-time constraints from the Baryon Acoustic Oscillations~(BAO) and integrated Sach-Wolfe~(iSW) effect. 

In Section \ref{basiceq}, a quintom model with two scalars is proposed as a realization of the extra component $X$.  A ``phantom''~(with negative kinetic energy term) scalar is assumed to couple to matter while the other scalar serves simply as the dark energy responsible for accelerated expansion of the Universe. The matter-phantom coupling stabilizes the phantom value by letting the matter dragging it along with the cosmological expansion.  The quintessence scalar serves as the DE with negligible contribution in the early times and only rise to dominate at the late epoch.  By tuning the model parameters and initial condition, a viable realistic quintom model is achieved that passes some of the early and late time constraints.  
%%%%%%%%%%%%%%%%%%%%%%%%%%%%%%%%%%%%%%%%%%%
%%%%%%%%%%%%%%%%%%%%%%%%%%%%%%%%%%%%%%%%%%%
%%%%%%%%%%%%%%%%%%%%%%%%%%%%%%%%%%%%%%%%%%%
%======================================%
%<<<<<<<<<<<< SECTION II  >>>>>>>>>>>>>>%
%======================================%
%%%%%%%%%%%%%%%%%%%%%%%%%%%%%%%%%%%%%%%%%%%
%%%%%%%%%%%%%%%%%%%%%%%%%%%%%%%%%%%%%%%%%%%
%%%%%%%%%%%%%%%%%%%%%%%%%%%%%%%%%%%%%%%%%%%

%%%%%%%%%%%%%%%%%%%%%%%%%%%%%%%%%%%%%%%%%%%
%%%%%%%%%%%%%%%%%%%%%%%%%%%%%%%%%%%%%%%%%%%
%%%%%%%%%%%%%%%%%%%%%%%%%%%%%%%%%%%%%%%%%%%
%%%%%%%%%%%%%%%%%%%%%%%%%%%%%%
\section{Quintom Dark Energy Model} \label{basiceq}
%%%%%%%%%%%%%%%%%%%%%%%%%%%%%%%%%%%%%%%%%%%%%%%%%%%%%%%%%%%%%%%%%%%%%%%%%%%%%%%%%%%%%%%%%%
We consider the quintom action with 2 scalar fields and interaction between matter fields and one of the scalar field as the following
\bea
S &=& \int d^4 x \sqrt{-g} \left[ \frac{1}{2\kappa^2} R - \frac{1}{2} (\partial \phi)^2 + \frac{1}{2}(\partial \sigma)^2 - V(\phi)\right] \nn 
& & + S_M (g_{\mu\nu}, \sigma, \psi_M) \,,
\ena
where $\kappa^2 = 8 \pi G$ is an inverse of the reduced Planck mass squared. R is the Ricci scalar, $\phi$ is a quintessence scalar field, $\sigma$ is a ``phantom'' scalar field, and $\psi_M$ is a matter field. We assume that there is only one self-interacting potential of the quintessence scalar field, while the ``phantom'' scalar field is rolling on an effective potential arising from the phantom-matter interaction as we will explain below. Strictly speaking, this $\sigma$ is not exactly the standard phantom but rather a ghost field since its equation of state is $P_{\sigma}=\rho_{\sigma}<0$.  However, here and henceforth we will simply call it the phantom field for convenience, and also in accordance with the original name quintom~(quintessence $+$ phantom). The extra component $X$ is identified with the phantom field $\sigma$ in this model. Note that the quintom model has negative kinetic energy term in the Lagrangian, the phantom scalar field thus encounters a quantum instability problem of its own. In our model, however, the {\it total} energy density of all components in the universe is always positive and the evolution of the universe always obey positive energy condition.

By varying the action with respect to $g^{\mu\nu}$, we obtain 
the equation of motion
\bea
R_{\mu\nu} - \frac{1}{2}g_{\mu\nu} R = \kappa^2 (T_{\mu\nu}^{(M)} + T_{\mu\nu}^{(\phi)} + T_{\mu\nu}^{(\sigma)}) \,,
\ena
where $T_{\mu\nu}^{(M)}$ is an energy-momentum tensor of the non-relativistic matter and radiation. Energy-momentum tensors of the quintessence scalar field and the phantom scalar field are given by
\bea
T_{\mu\nu}^{(\phi)} &=& \partial_{\mu} \phi \partial_{\nu} \phi + g_{\mu\nu} \left(- \frac{1}{2} (\partial \phi)^2 - V(\phi)\right) \,, \\
T_{\mu\nu}^{(\sigma)} &=& - \partial_{\mu} \sigma \partial_{\nu} \sigma + g_{\mu\nu} \left(\frac{1}{2} (\partial \sigma)^2 \right) \,,
\ena
respectively. Using the flat Friedmann-Lema\^{\i}tre-Robertson-Walker (FLRW) metric, $ds^2 = - dt^2 + a^2 (t)d \bf{x}^2$, and assuming $\phi = \phi(t)$ and $\sigma = \sigma(t)$, we obtain the Friedmann equations,
\bea
3H^2 &=& \kappa^2 (\rho_m + \rho_r + \rho_{\phi} + \rho_{\sigma}) \,, \label{fm1}\\
3H^2 + 2\dot H &=& \kappa^2 (-P_m -P_r - P_{\phi} - P_{\sigma}) \,. \label{fm2}
\ena
$\rho_m$, $\rho_r$, $P_m$, and $P_r$ are energy densities and pressures of non-relativistic matter and radiation, respectively. The notation ``$ ~ ^. ~ $'' means a derivative with respect to time. The energy densities and pressures of the scalar fields are  
\bea
\rho_{\phi} &=& \frac{1}{2}\dot \phi^2 + V(\phi) \,, ~~~  \rho_{\sigma} = -\frac{1}{2}\dot \sigma^2 \,, \\
P_{\phi} &=& \frac{1}{2}\dot \phi^2 - V(\phi) \,, ~~~ P_{\sigma} = -\frac{1}{2}\dot \sigma^2 \,.
\ena
We then can define an equation of state parameter of the dark energy and an effective equation of state parameter as
\bea
w_{DE} &\equiv& \frac{P_{\phi} + P_{\sigma}}{\rho_{\phi} + \rho_{\sigma}} = \frac{\frac{1}{2}\dot \phi^2 -\frac{1}{2}\dot \sigma^2 - V(\phi)}{\frac{1}{2}\dot \phi^2 -\frac{1}{2}\dot \sigma^2 + V(\phi)} \,, \\
w_{\rm eff} &\equiv& \frac{P_{\phi} + P_{\sigma} + P_m + P_r}{\rho_{\phi} + \rho_{\sigma} + \rho_m + \rho_r} = -1 - \frac{2}{3}\frac{\dot H}{H^2} \,.
\ena
Remark that the $w_{DE}$ in this section and henceforth is contributed by both scalar fields, which is different from the $w_{DE}$ in the Section \ref{GenPhen}.
For each scalar field, their equation of state parameters are
\bea
w_{\phi} = \frac{P_{\phi}}{\rho_{\phi}} = \frac{\frac{1}{2}\dot \phi^2 - V(\phi)}{\frac{1}{2}\dot \phi^2 + V(\phi)} \,, ~~~ w_{\sigma} = \frac{P_{\sigma}}{\rho_{\sigma}} = +1 \,.
\ena
Note that $\rho_{\sigma}$ is negative, and $w_{\sigma}$ is always equal to $+1$. These are crucial in resolving the Hubble tension problem which we will show in the Section \ref{dynamical}.

%Taking covariant derivative on the field equation we find
%\bea
%0 = \nabla_{\mu} T^{\mu (M)}_{\nu} + \nabla_{\mu} T^{\mu (\phi)}_{\nu} + \nabla_{\mu} T^{\mu (\sigma)}_{\nu} \,.
%\ena
We assume there is only an interaction between matter field and the phantom field, i.e. $\nabla_{\mu} T^{\mu (\phi)}_{\nu} = 0$. In this work we consider the interaction in the form
\bea
\nabla_{\mu} T^{\mu (M)}_{\nu} &=& \kappa \delta T_M \nabla_{\nu}\sigma \,, \\ 
\nabla_{\mu} T^{\mu (\sigma)}_{\nu} &=& -\kappa \delta T_M \nabla_{\nu}\sigma \,,
\ena
where $T_M = -\rho_M + 3P_M$, and $\delta$ is a dimensionless constant. This is a conformal interaction form which arises in many scalar-tensor theories after taking a conformal transformation to the Einstein frame \cite{Amendola:1999er,Tsujikawa:2010sc}. Hence, the continuity equations are
\bea
\dot \rho_m + 3H \rho_m &=& \kappa \delta \rho_m \dot \sigma \,, \label{continuitymatter} \\
\dot \rho_{\sigma} + 3H (\rho_{\sigma} + P_{\sigma}) &=& - \kappa \delta \rho_m \dot \sigma \,, \label{continuitysig}\\
\dot \rho_{\phi} + 3H (\rho_{\phi} + P_{\phi}) &=& 0 \,, \label{continuityphi}\\
\dot \rho_r + 4H \rho_r &=& 0 \,.
\ena
Substituting energy density and pressure of each scalar field we find the equations of motion
\bea
\ddot \sigma + 3H \dot \sigma = \kappa \delta \rho_m \,, \\
\ddot \phi + 3H \dot \phi + V_{,\phi} = 0 \,.
\ena
The right-hand-side~(RHS) of the equation of motion of the phantom scalar field acts as an effective potential. This is similar to the effective potential in the chameleon or symmetron gravity \cite{Khoury:2003aq,Khoury:2003rn,Hinterbichler:2011ca}

%%%%%%%%%%%%%%%%%%%%%%%%%%%%%%%%%%%%%%%%%%%
%%%%%%%%%%%%%%%%%%%%%%%%%%%%%%%%%%%%%%%%%%%
%%%%%%%%%%%%%%%%%%%%%%%%%%%%%%%%%%%%%%%%%%%
%======================================%
%<<<<<<<<<<<< SECTION III  >>>>>>>>>>>>>>%
%======================================%
%%%%%%%%%%%%%%%%%%%%%%%%%%%%%%%%%%%%%%%%%%%
%%%%%%%%%%%%%%%%%%%%%%%%%%%%%%%%%%%%%%%%%%%

\section{Dynamical System}
\label{dynamical}
%%%%%%%%%%%%%%%%%%%%%%%%%%%%%%%%%%%%%%%%%%%
%%%%%%%%%%%%%%%%%%%%%%%%%%%%%%%%%%%%%%%%%%%
%%%%%%%%%%%%%%%%%%%%%%%%%%%%%%%%%%%%%%%%%%%

\subsection{Autonomous Equations and Fixed Points}
We will use a dynamical system approach to study cosmological scenarios of the quintom dark energy model through the behaviour of their fixed points. First, the dimensionless dynamical variables are defined as the following
\bea
x_1 \equiv \frac{\kappa \dot\phi}{\sqrt{6}H} \,, ~ x_2 \equiv \frac{\kappa \sqrt{V(\phi)}}{\sqrt{3}H} \,, ~ x_3 \equiv \frac{\kappa \dot\sigma}{\sqrt{6}H} \,, ~ x_4 \equiv \frac{\kappa \sqrt{\rho_r}}{\sqrt{3}H} \,. \nn
\ena
According to the Friedmann equation (\ref{fm1}), the density parameters in terms of the dynamical variables are
\bea
\Omega_m &=& 1-x_1^2 -x_2^2 + x_3^2 -x_4^2 \,, \label{omegam} \\
\Omega_r &=& x_4^2 \,, \\
\Omega_{DE} &=& x_1^2 + x_2^2 - x_3^2 \,, \\
\Omega_{\phi} &=& x_1^2 + x_2^2 \,, \\
 \Omega_{\sigma} &=& - x_3^2 \,, 
\ena
where $\Omega_m \equiv \kappa^2 \rho_m / 3H^2$. In addition, the equation of states are
\bea
w_{DE} &=& \frac{x_1^2 -x_2^2 - x_3^2}{x_1^2 + x_2^2 - x_3^2} \,, \\
w_{\phi} &=& \frac{x_1^2 -x_2^2}{x_1^2 + x_2^2} \,, \\ 
w_{\sigma} &=& 1 \,, \\
w_{\rm eff} &=& x_1^2 - x_2^2 - x_3^2 + \frac{x_4^2}{3} \label{weff} \,. ~~
\ena
In the last equation we have used the second Friedmann equation (\ref{fm2}) which leads to
\bea
\frac{\dot H}{H^2} = - \frac{1}{2}\left(3 + 3x_1^2 - 3x_2^2 - 3x_3^2 + x_4^2 \right) \,. \label{hubbleev}
\ena
Differentiating the dynamical variables with respect to $N$, where $N =\ln a$ is an e-folding number, we find a set of autonomous equations:
\bea
\frac{dx_1}{dN} &=& \frac{\sqrt{6}}{2} \lambda_{\phi} x_2^2 - 3x_1 - x_1 \frac{\dot H}{H^2} \,, \label{auto1} \\
\frac{dx_2}{dN} &=& - \frac{\sqrt{6}}{2} \lambda_{\phi} x_1 x_2 - x_2 \frac{\dot H}{H^2} \,, \label{auto2} \\
\frac{dx_3}{dN} &=& \frac{\sqrt{6}}{2}\delta (1-x_1^2 -x_2^2 + x_3^2 -x_4^2) \nn
& & - 3x_3 - x_3 \frac{\dot H}{H^2} \,, \label{auto3} \\
\frac{dx_4}{dN} &=& -2x_4 - x_4 \frac{\dot H}{H^2} \,, \label{auto4}
\ena
where $\lambda_{\phi} \equiv - V_{,\phi} / \kappa V$. For an exponential form of a potential, i.e. $V(\phi) = V_0 e^{- \kappa \lambda_{\phi} \phi}$, $\lambda_{\phi}$ is a constant. Then the autonomous equations are closed. 

Fixed points of the system can be obtained by setting $dx_1/dN = dx_2/dN = dx_3/dN = dx_4/dN = 0$. They are shown in %Table \ref{fixedpoints} 
Appendix \ref{fixedpoint}. The dynamical variables $x_2$ and $x_4$ are always positive, while $x_1$ and $x_3$ can be positive or negative depending on the signs of $\dot\phi$ or $\dot\sigma$. We are interested only in the case where $\lambda_{\phi} > 0$ (an exponential decay) and $\delta > 0$. With these fixed points, density parameters and equation of state parameters are represented in Table \ref{densityandequationofstate}.

\begin{widetext}

\begin{table}[thb]
\begin{center}
  \begin{tabular}{|c||c|c|c|c|c|}
\hline 
& & & & & \\[-.5em]
&$\Omega_m$&$\Omega_r$&$\Omega_{DE}$&$w_{DE}$&$w_{\rm eff}$
\\[.5em]
\hline
& & & & & \\[-.5em]
(a)&$0$&$0$&1&$1$&$1$
\\[.5em]
\hline
& & & & & \\[-.5em]
(b)&$0$&$1$&$0$&$-$&$\frac{1}{3}$
\\[.5em]
\hline
& & & & & \\[-.5em]
(c)&$-\frac{1}{3\delta^2}$&$1+\frac{1}{2\delta^2}$ &$- \frac{1}{6\delta^2}$&$1$&$\frac{1}{3}$
\\[.5em]
\hline
& & & & & \\[-.5em]
(d)&$1+ \frac{2\delta^2}{3}$&$0$&$- \frac{2\delta^2}{3}$&$1$&$- \frac{2\delta^2}{3}$
\\[.5em]
\hline
& & & & & \\[-.5em]
(e)&$0$&$1 - \frac{4}{\lambda_{\phi}^2}$&$\frac{4}{\lambda_{\phi}^2}$&$\frac{1}{3}$&$\frac{1}{3}$
\\[.5em]
\hline
& & & & & \\[-.5em]
(f)&$- \frac{1}{3\delta^2}$&$1 + \frac{1}{2\delta^2} - \frac{4}{\lambda_{\phi}^2}$&$- \frac{1}{6\delta^2} + \frac{4}{\lambda_{\phi}^2}$&$\frac{8\delta^2 - \lambda_{\phi}^2}{24\delta^2 - \lambda_{\phi}^2}$&$\frac{1}{3}$
\\[.5em]
\hline
& & & & & \\[-.5em]
(g)&$0$&$0$&$1$&$-1 + \frac{\lambda_{\phi}^2}{3}$&$-1 + \frac{\lambda_{\phi}^2}{3}$
\\[.5em]
\hline
& & & & & \\[-.5em]
(h)&$\frac{(\lambda_{\phi}^2 -3)(3\lambda_{\phi}^2 + 2\delta^2 (\lambda_{\phi}^2 - 6))}{3(\lambda_{\phi}^2 - 2\delta^2)^2}$&$0$&$\frac{12\delta^4 + 9\lambda_{\phi}^2 - 2\delta^2 (18 - 3 \lambda_{\phi}^2 + \lambda_{\phi}^4)}{3(\lambda_{\phi}^2 - 2\delta^2)^2}$&$\frac{2\delta^2 (2\delta^2 - \lambda_{\phi}^2)(\lambda_{\phi}^2 - 3)}{12\delta^4 + 9 \lambda_{\phi}^2 - 2 \delta^2 (18 - 3\lambda_{\phi}^2 + \lambda_{\phi}^4)}$&$\frac{2\delta^2 (\lambda_{\phi}^2 - 3)}{6\delta^2 - 3\lambda_{\phi}^2}$
\\[.5em]
\hline       

 \end{tabular}
    \caption{Density parameters and equation of state parameters of each fixed point.}
\label{densityandequationofstate}
\end{center}
\end{table}

\end{widetext}

In the next section, stability of each fixed point will be examined by considering their corresponding eigenvalues.

%%%%%%%%%%%%%%%%%%%%%%%%%%%%%%%%%%%%%%%%%%%%%%%%%%%%%%%%%%%%%%%%%%%%%%%%%%%%%%%%%%%%%%%%%%%%%
\subsection{Eigenvalues of fixed points}

Eigenvalues of each fixed point in Table \ref{fixedpoints} are as the following (definition of eigenvalues is represented in Appendix \ref{matrixperturbations}).

\subsubsection{Fixed Point (a)}

Eigenvalues of the fixed point are
\bea
\mu_{(a)} = 1\,, 0\,, 3\pm \sqrt{6} \, \delta \sqrt{x_1^2-1}\,, 3-\sqrt{\frac{3}{2}} \lambda_{\phi}  x_1 \,.
\ena
Although a sign $\pm$ depends on roots of the condition $x_1^2 - x_3^2 = 1$, the fixed point is either saddle or unstable point. Since this fixed point does not match with any known cosmological era, we no longer consider it.  

\begin{widetext}
\subsubsection{Radiation Dominated Solutions}

Eigenvalues of the fixed point (b), (c), (e), and (f) are given by
\bea
\mu_{(b)} &=& 2 \,, -1 \,, -1 \,, 1 \,. \\
\mu_{(c)} &=& -1 \,, 2 \,, -\frac{1}{2} \pm \frac{1}{2}\sqrt{- \left( \frac{2}{\delta^2} + 3\right)} \,, \\
\mu_{(e)} &=& -1 \,, 1 \,, - \frac{1}{2} \pm \sqrt{\frac{16}{\lambda_{\phi}^2} - \frac{15}{4}} \,, \\
\mu_{(f)} &=& -\frac{1}{2} \pm \frac{1}{2\delta^2 \lambda_{\phi}^2} \sqrt{-\delta^2 \lambda_{\phi}^4 + \delta^4 (32 \lambda_{\phi}^2 - 9 \lambda_{\phi}^{4}) + \sqrt{\delta^4 \lambda_{\phi}^4 (\lambda_{\phi}^4 + 4 \delta^4 (16 - 3 \lambda_{\phi}^2)^2 - 4\delta^2 \lambda_{\phi}^2 (16 + 3\lambda_{\phi}^2))}} \,, \\ 
& &  -\frac{1}{2} \pm \frac{1}{2\delta^2 \lambda_{\phi}^2} \sqrt{-\delta^2 \lambda_{\phi}^4 + \delta^4 (32 \lambda_{\phi}^2 - 9 \lambda_{\phi}^{4}) - \sqrt{\delta^4 \lambda_{\phi}^4 (\lambda_{\phi}^4 + 4 \delta^4 (16 - 3 \lambda_{\phi}^2)^2 - 4\delta^2 \lambda_{\phi}^2 (16 + 3\lambda_{\phi}^2))}} \,.
\ena
Therefore, the fixed point (b), (c), and (e) are saddle points. For the point (f), we can understand the behaviour of the fixed point when we set the value of $\lambda_{\phi}$ and $\delta$.

\subsubsection{Matter Dominated Solutions}

For the fixed point (d) and (h), their corresponding eigenvalues are
\bea
\mu_{(d)} &=& - \frac{3}{2} - \delta^2 \,, - \frac{3}{2} - \delta^2 \,, - \frac{1}{2} - \delta^2 \,, \frac{3}{2} - \delta^2 \,, \\
\mu_{(h)} &=& \frac{\lambda_{\phi}^2 + 2\delta^2 (\lambda_{\phi}^2 - 4)}{4\delta^2 - 2 \lambda_{\phi}^2} \,,  \frac{3\lambda_{\phi}^2 + 2\delta^2 (\lambda_{\phi}^2 - 6)}{4\delta^2 - 2 \lambda_{\phi}^2} \,, \nn 
& & \frac{1}{4(\lambda_{\phi}^2 - 2\delta^2)^2} \left(-3\lambda_{\phi}^4 - 2\delta^2 \lambda_{\phi}^2 (\lambda_{\phi}^2 - 9) + 4 \delta^4 (\lambda_{\phi}^2 - 6) \right.\nn
& & \left. \pm \sqrt{3(\lambda_{\phi}^2 - 2 \delta^2)^2(72 \lambda_{\phi}^2 - 21 \lambda_{\phi}^4 + 4\delta^2 (-72 + 18 \lambda_{\phi}^2 + \lambda_{\phi}^4) + 4\delta^4 (60 - 28 \lambda_{\phi}^2 + 3\lambda_{\phi}^4))}\right) \,.
\ena
\end{widetext}
The fixed point (d) is stable when $\delta^2 > \frac{3}{2}$, whereas it is a saddle point when $\delta^2 < \frac{3}{2}$. For the fixed point (h), the eigenvalues can be understood once we set the value of $\lambda_{\phi}$ and $\delta$.

\subsubsection{Accelerated Expansion Solutions}

Eigenvalues of the fixed point (g) are
\bea
\mu_{(g)} = \frac{1}{2} (\lambda_{\phi}^2 - 6) \,, \frac{1}{2} (\lambda_{\phi}^2 - 6)\,, \frac{1}{2} (\lambda_{\phi}^2 - 4) \,, \lambda_{\phi}^2 - 3 \,. \nn
\ena
Thus, the fixed point is stable when $\lambda_{\phi}^2 < 3$. For the point (h) it is the same as the previous case.

%======================================%
%<<<<<<<<<<<< SECTION IV  >>>>>>>>>>>>>>%
%======================================%
%%%%%%%%%%%%%%%%%%%%%%%%%%%%%%%%%%%%%%%%%%%%%%%%%%%
%%%%%%%%%%%%%%%%%%%%%%%%%%%%%%%%%%%%%%%%%%%%%%%%%%%
%%%%%%%%%%%%%%%%%%%%%%%%%%%%%%%%%%%%%%%%%%%%%%%%%%%

\section{Numerical Solutions}
\label{numer}
%%%%%%%%%%%%%%%%%%%%%%%%%%%%%%%%%%%%%%%%%%%%%%%%%%%
%%%%%%%%%%%%%%%%%%%%%%%%%%%%%%%%%%%%%%%%%%%%%%%%%%%

\begin{figure}[h]
\includegraphics[width=7.5cm]{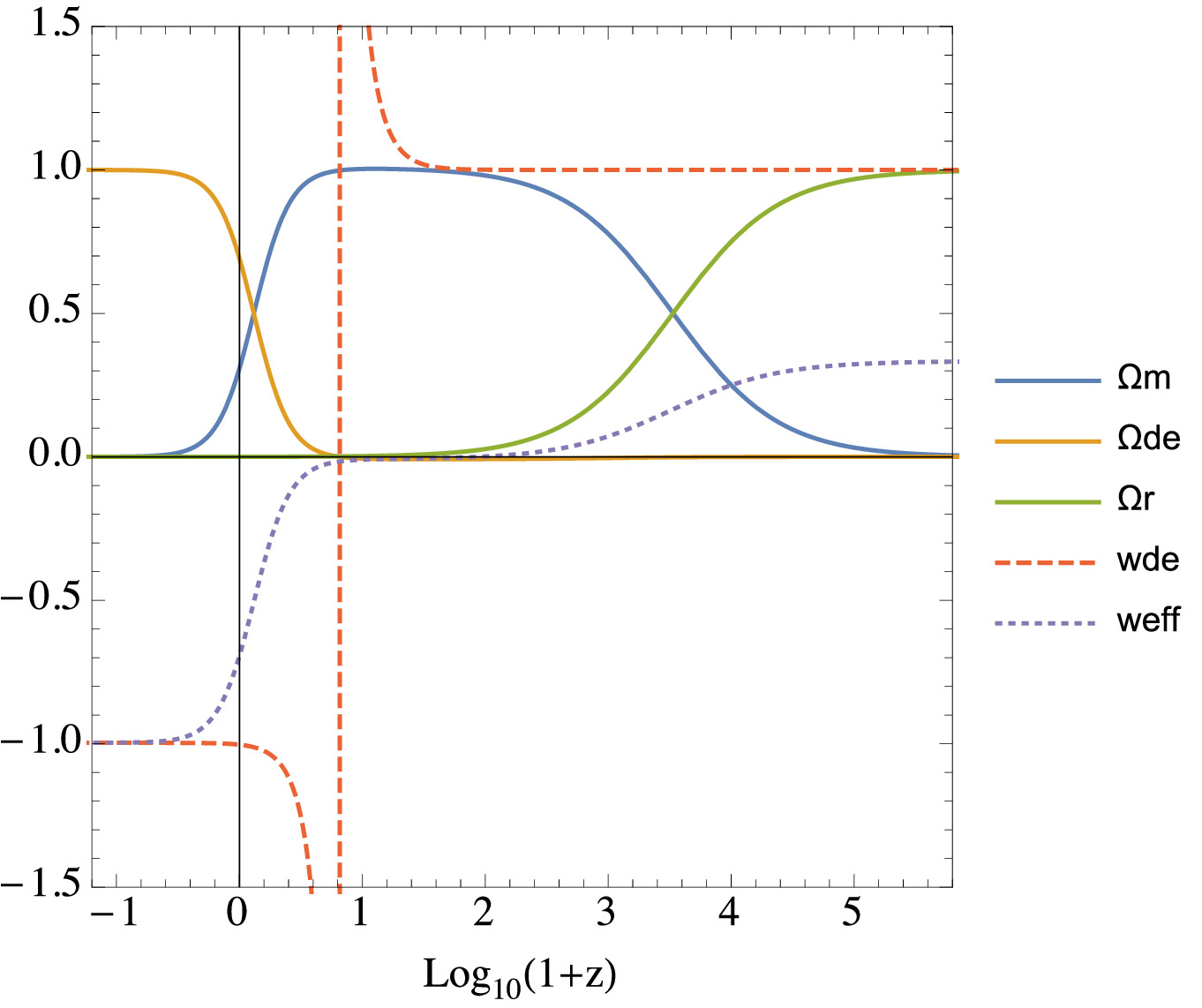}
\\
\includegraphics[width=7.5cm]{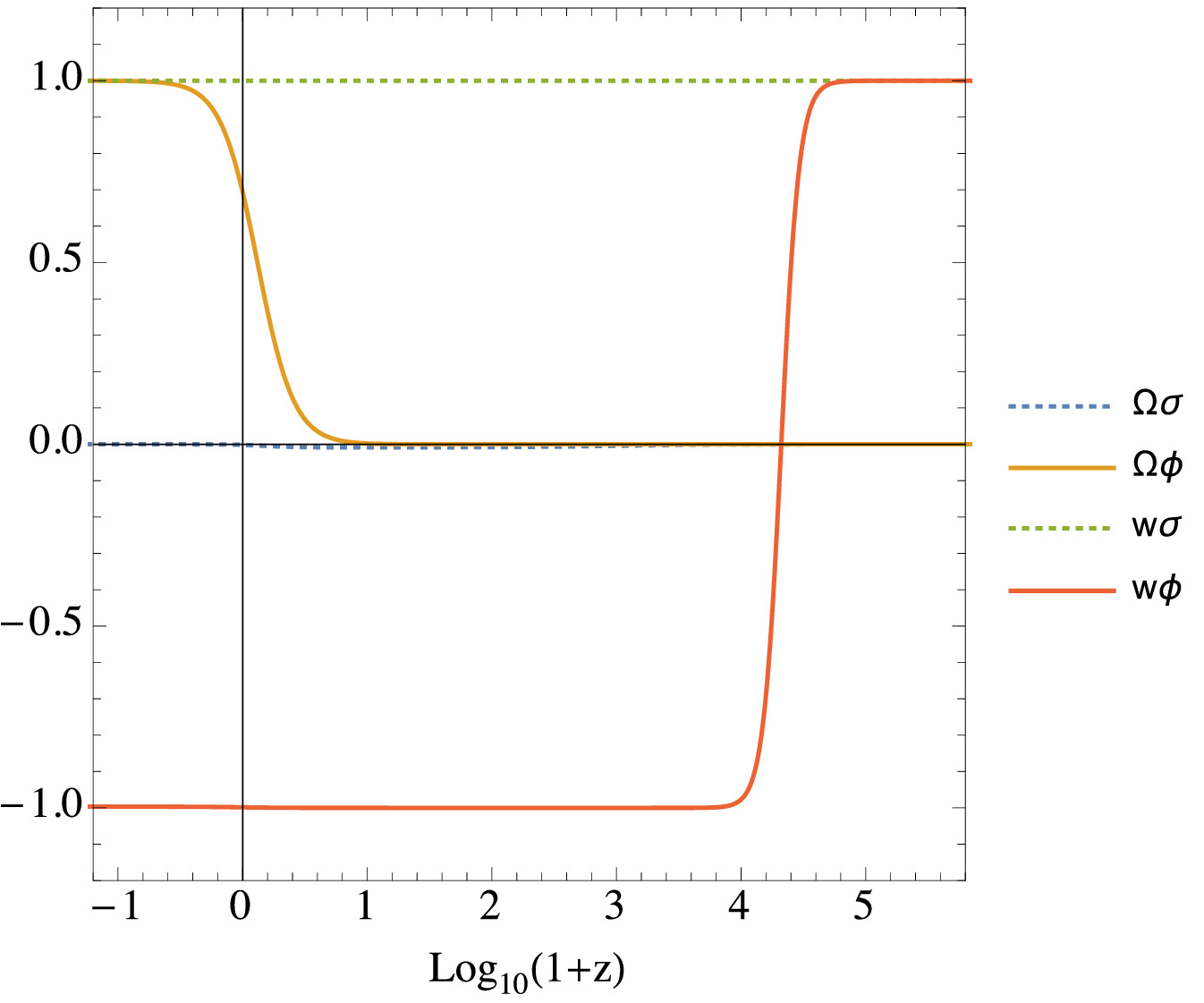}
\\
\includegraphics[width=7.7cm]{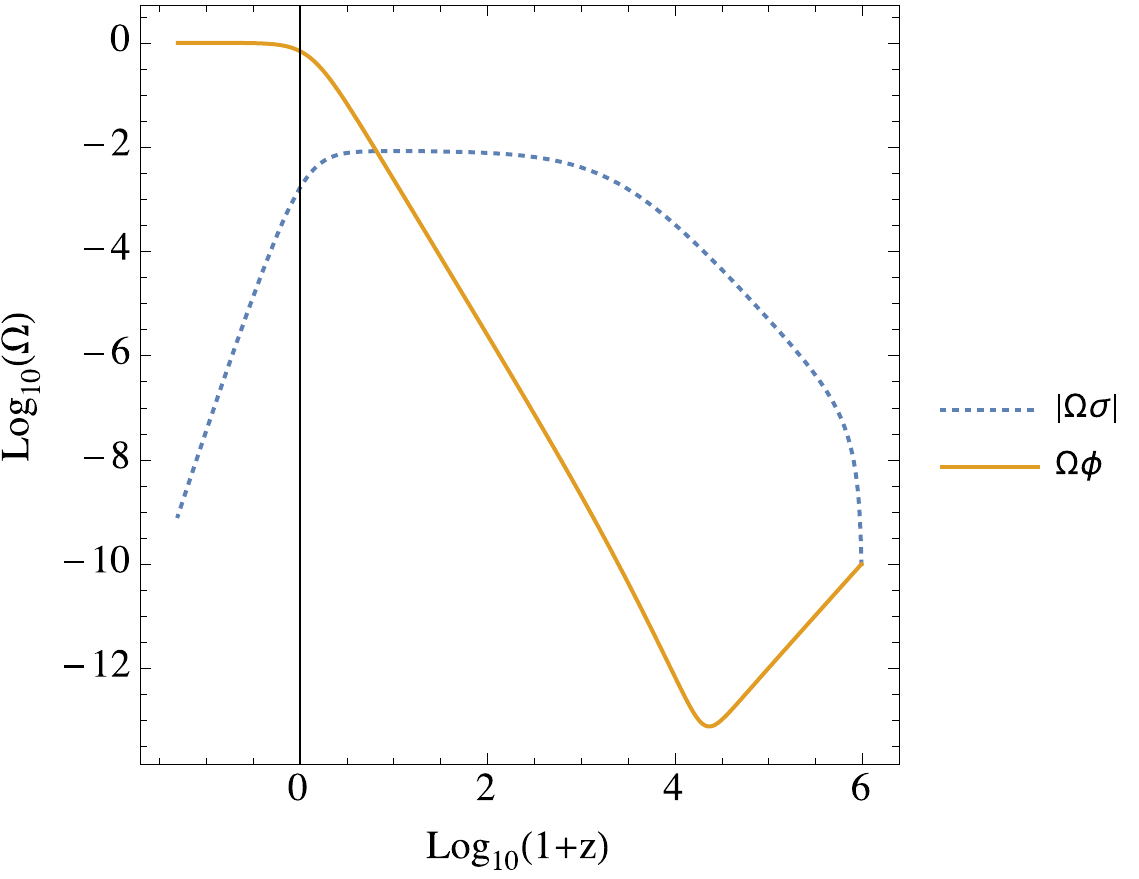}
\caption{Evolution of the density parameters and the equation of state parameters according to Eqs. (\ref{omegam}) - (\ref{weff}) where initial conditions are $x_1 = 1 \times 10^{-5}$, $x_2 = 1 \times 10^{-10}$, $x_3 = 1 \times 10^{-5}$, and $x_4 = 0.9983$ at $z = 9.74811 \times 10^5~(N= -13.79)$.}
\label{evolution}
\end{figure}

In this section, the autonomous equations (\ref{auto1}) - (\ref{auto4}) are solved numerically, where we set $\lambda_{\phi} = 0.1$ and $\delta = 0.113$. This choice lies within the range of values satisfying the fixed-points scenario $\lambda_{\phi}<\sqrt{3},\delta<\sqrt{3/2}$ discussed previously and the condition $w_{\rm eff}<-1/3$ at the fixed point (g) which gives $\lambda_{\phi}<\sqrt{2}$ for accelerated expansion. By tuning the model parameters and initial condition, we can obtain the Hubble parameter in the desired range of values, namely $H_{0} \approx 74$ km s$^{-1}$ Mpc$^{-1}$ in order to relieve the Hubble tension. Since the viable cosmology requires sufficiently long period of radiation dominated era, we choose the initial point of numerical solution to be deep into the radiation epoch at $z=9.74811 \times 10^5$ corresponding to $N= -\ln(1+z)= -13.79$ in order to guarantee long radiation epoch. The cosmological evolution is actually insensitive to the choice of the initial value of $N$, e.g. setting $N=-13.81$ would give indistinguishable results as long as we define the present-day $N_{0}$ such that $\Omega_{m}^{(0)}$ has the same value. When we change the initial value of $N$, the present-day value $N_{0}$ at some fixed choice of $(\Omega_{m}^{(0)},\Omega_{r}^{(0)})$ will simply shift accordingly. The overall cosmological evolution remains the same with the physical redshift defined as $N-N_{0}=-\ln(1+z)$.  Evolution of the density parameters and the equation of state parameters are shown in Fig. \ref{evolution}.

Figure \ref{evolution} demonstrates a viable cosmological scenario where the Universe evolved from the radiation dominated era to the matter dominated epoch, and followed by the late-time accelerated expansion. Since $\lambda_{\phi} = 0.1$ and $\delta = 0.113$, the fixed points (e), (f), and (h) do not exist, while the fixed point (c) yields $\Omega_r \approx 40$ which is too large. Since we are interested in $\Omega_r \approx 1$, we choose initial condition closed to the fixed point (b). The matter dominated era is point (d) automatically, and the accelerated expansion is the point (g). Therefore, the cosmological viable evolution is
\bea
{\rm (b)} ~~ \rightarrow ~~ {\rm (d)} ~~ \rightarrow ~~ {\rm (g)} \,.
\ena

\begin{figure}[h]
~~~~~~~~~~~\includegraphics[width=7.5cm]{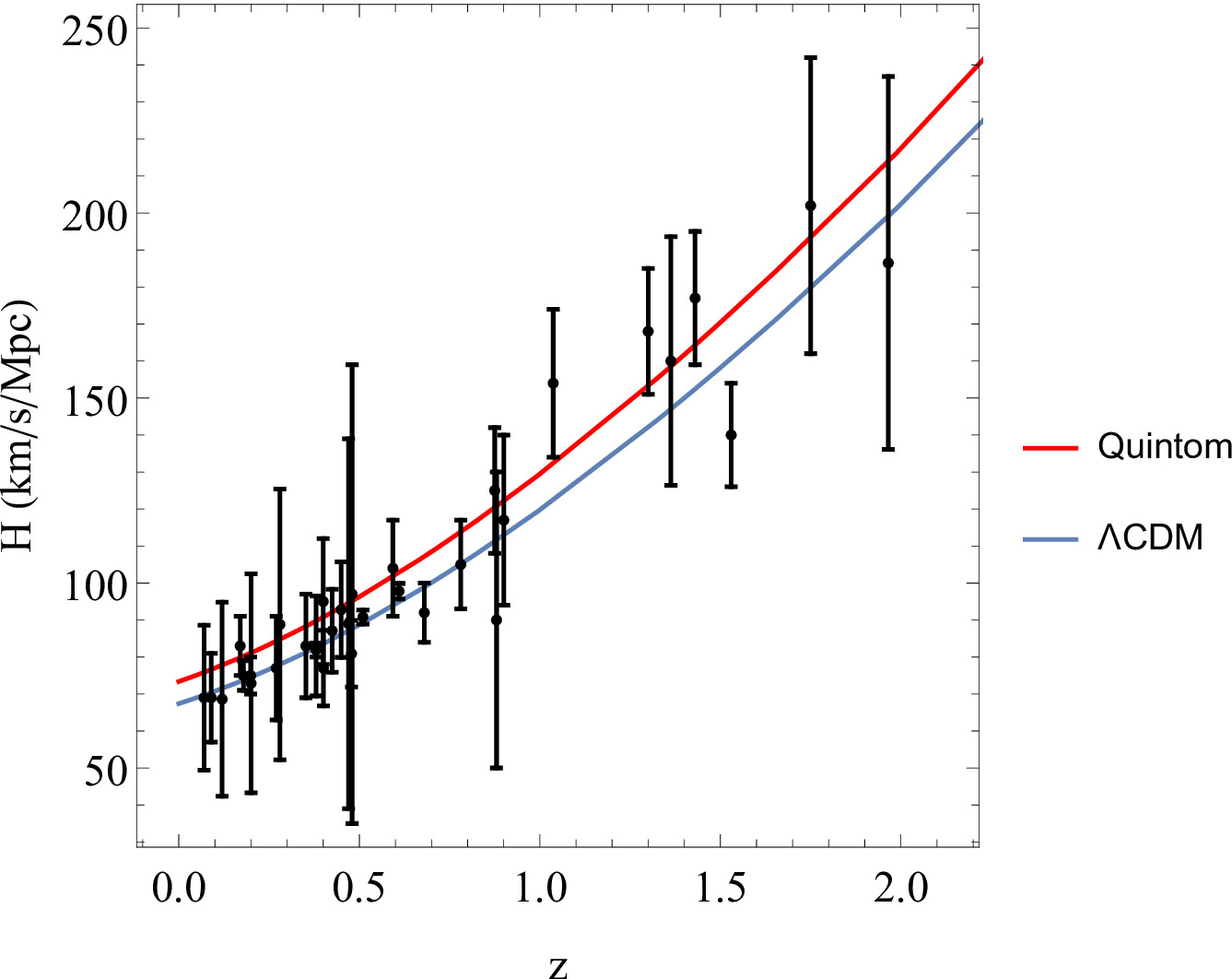}
\\
\includegraphics[width=5.5cm]{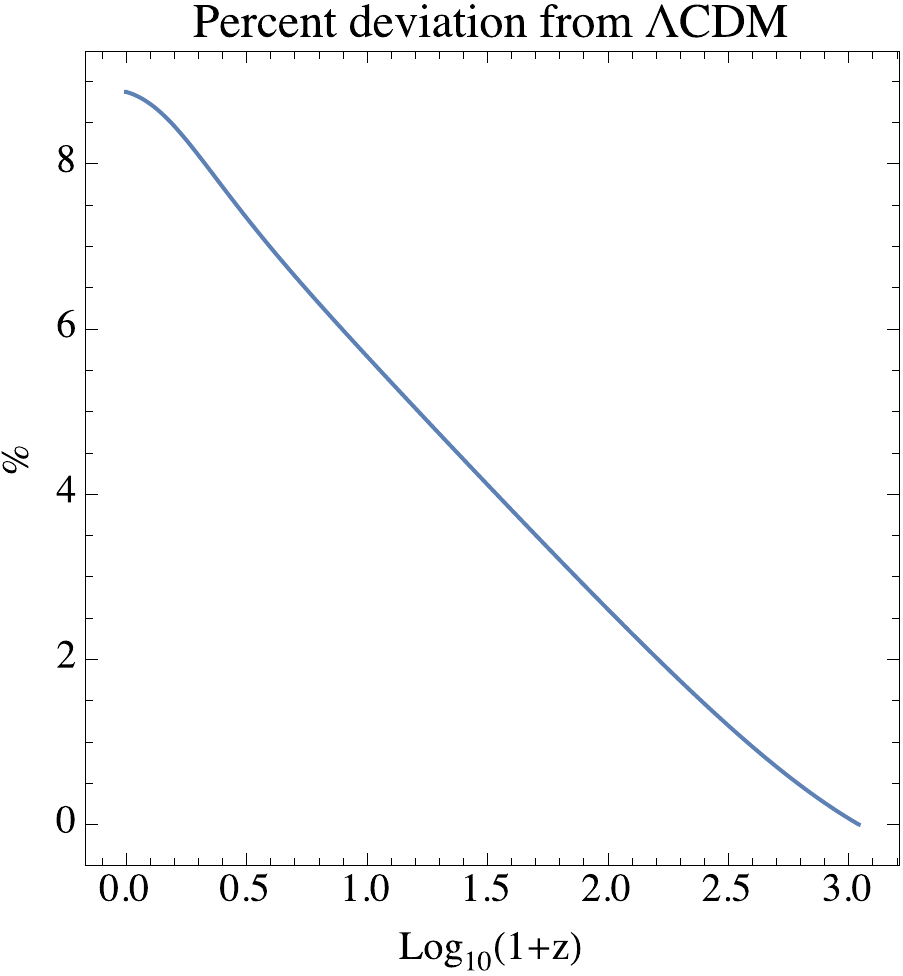}
\caption{Top figure represents evolutions of the Hubble parameters at late-time, while the bottom figure represents percent deviation from $\Lambda$CDM from the last-scattering surface to the present.}
\label{hubbleevo}
\end{figure}

According to the middle figure of Fig.~\ref{evolution}, the density of the phantom scalar field is negative and $w_{\sigma} = 1$ as desired, while the density of dark energy (quintessence + phantom) increases at late-time. 
We can obtain the evolution of the Hubble parameter by integrating Eq. (\ref{hubbleev}),
\bea  \label{HNeq}
H(N) &=& C \exp\left[\frac{1}{2} \left(-3 N - 3 \int x_1^2 dN + 3\int x_2^2 dN \right.\right. \nn
& & \left.\left. + 3 \int x_3^2 dN - \int x_4^2 dN\right)\right] \,,
\ena 
where $C$ is a constant of integration which can be obtained by comparing the above equation to the Hubble parameter from the $\Lambda$CDM at the last-scattering surface. $x_1$, $x_2$, $x_3$, and $x_4$ are obtained from numerical solutions with the same initial conditions used in Fig.~\ref{evolution}. We set $H_0 = 67.4~{\rm km ~s}^{-1}{\rm Mpc}^{-1}$ to find the Hubble parameter at $z = 1100$ with the $\Lambda$CDM model, and then we start the evolution in the quintom model from this value of $H(1100)$. The evolution of Hubble parameter of the quintom compared to that of the $\Lambda$CDM is shown in Fig.~\ref{hubbleevo}.

In Fig. \ref{hubbleevo}, the Hubble parameter of the quintom model decreases at slightly different rate comparing to the $\Lambda$CDM, where we find $H_0 = 73.356~ {\rm km ~s}^{-1}{\rm Mpc}^{-1}$ at the present time. The data of $H(z)$ is from Ref.~\cite{Yu:2017iju}. Remarkably, the Hubble tension is alleviated. Note that the value of $H_0$ depends on the initial conditions and the values of $\lambda_{\phi}$ and $\delta$ which can be tuned to provide better precision comparing to the observations as will be shown subsequently. 

Cosmological parameters at the present time obtained from numerical simulations are represented in Table \ref{presentvalues}. These parameters correspond to the redshift at $z = 0$ in Figs \ref{evolution} and \ref{hubbleevo}.

\begin{table}[ht]
\begin{center}
  \begin{tabular}{|c|c|c|c|c|}
\hline
& & & & \\[-.5em]
$\Omega_m^{(0)}$&$\Omega_{DE}^{(0)}$&$\Omega_r^{(0)}$&$\Omega_{\sigma}^{(0)}$&$\Omega_{\phi}^{(0)}$
\\[.5em]
\hline
& & & & \\[-.5em]
$0.3078$&$0.69211$&$7.66\times 10^{-5}$&$-0.00164$&$0.69376$
\\[.5em]
\hline
\hline
& & & & \\[-.5em]
$w_{DE}$&$w_{\rm eff}$&$w_{\sigma}$&$w_{\phi}$&
\\[.5em]
\hline 
& & & & \\[-.5em]
$-1.003$&$-0.694$&$1$&$-0.9985$&$$
\\[.5em]
\hline

 \end{tabular}
    \caption{Cosmological parameters at the present time from the quintom model.}
\label{presentvalues}
\end{center}
\end{table}

\begin{figure}[h]
\includegraphics[width=8.2cm]{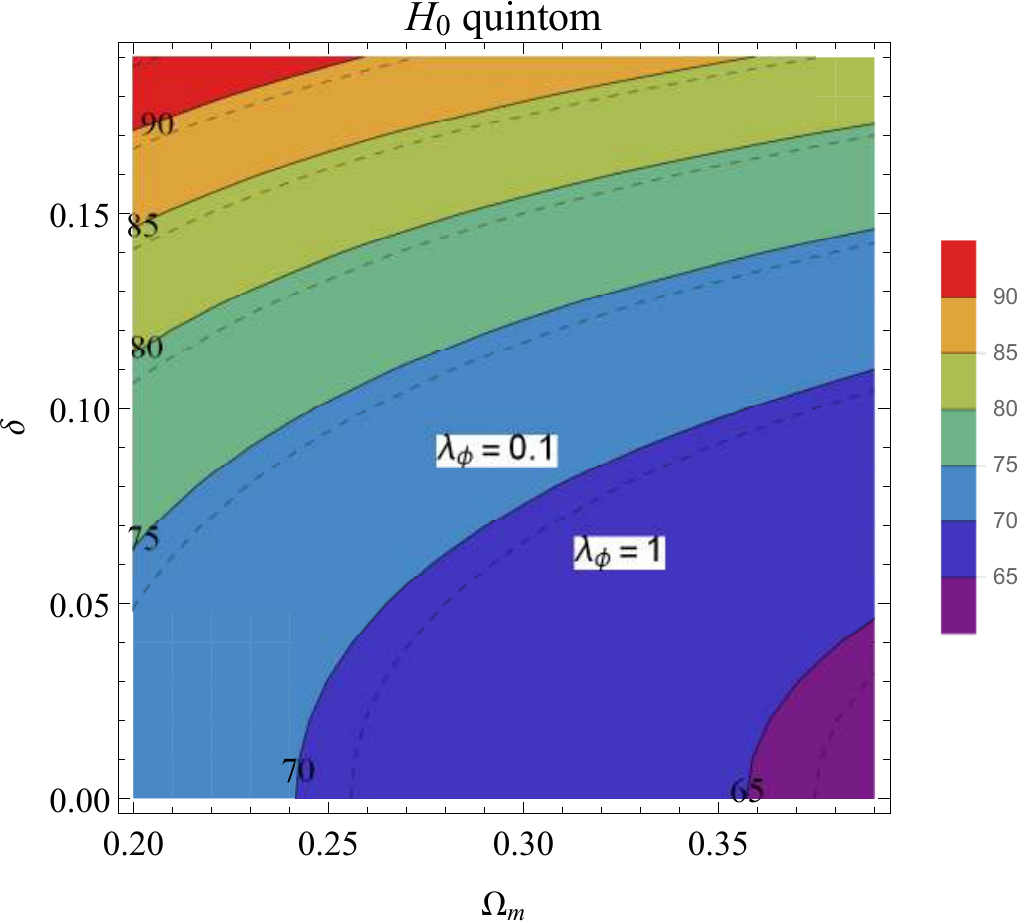}

\caption{Parameter space of the coupled quintom model for fixed $\lambda_{\phi}$, the number on each pairs of contour is the value of the corresponding $H_{0}$. The solid~(dashed) line represents contour with $\lambda_{\phi}=0.1~(1)$ respectively. The parameter space changes only slightly with $\lambda_{\phi}$.}
\label{H0para1}
\end{figure}

The motivation of this work is to find a modification to the standard $\Lambda$CDM model in such a way that the early-time parameters are mostly unchanged while additional phantom field only affects a small but accumulative significant change in the value of $H_{0}$. We thus explore the parameter space of the coupled quintom model with respect to $H_{0}$, where $\lambda_{\phi}<\sqrt{2}, \delta<\sqrt{3/2}$ required for a late-time accelerated expansion. In Fig.~\ref{H0para1}, the contour of constant $H_{0}$ is shown with respect to the phantom-matter coupling $\delta$ and the present-day matter density parameter $\Omega_{m}^{(0)}$ where the superscript is suppressed. The value of $H_{0}$ depends not only on $\Omega_{m}^{(0)}$ but also on the coupling $\delta$. Notably, the present-day Hubble parameter depends quite weakly on $\lambda_{\phi}$ which governs the quintessence evolution.   

\begin{figure}[h]
\includegraphics[width=8.4cm]{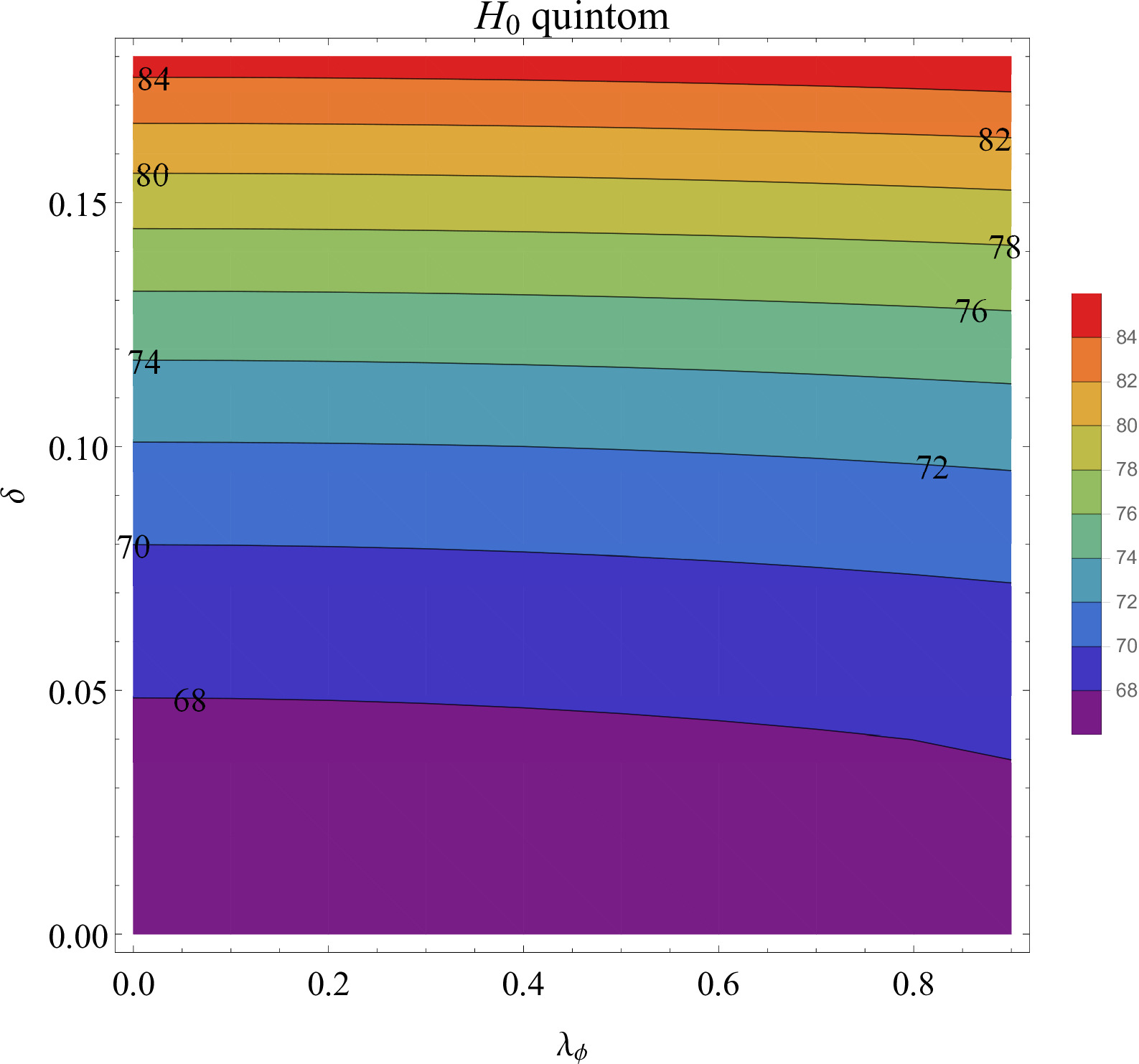}

\caption{Parameter space of the coupled quintom model under the condition $\Omega_{m}^{(0)}=0.308,\Omega_{\gamma}^{(0)}=5.38\times 10^{-5}, N_{\rm eff}=3.13$, the number on each contour is the value of the corresponding $H_{0}$. Region with small $(\lambda_{\phi}, \delta)$ gives $H_{0}\simeq 67-68~{\rm km ~s}^{-1}{\rm Mpc}^{-1}$, close to the value of the $\Lambda$CDM model. }
\label{H0para}
\end{figure}

To see the viability of the coupled quintom model, we plot the contour of each value of $H_{0}$ in the parameter space with respect to the constraints $\Omega_{m}^{(0)}=0.308,\Omega_{\gamma}^{(0)}=5.38\times 10^{-5}, N_{\rm eff}=3.13$ in Fig.~\ref{H0para}. The value of $H_{0}$ varies significantly with $\delta$ but relatively insensitive to the parameter $\lambda_{\phi}$. For a given value of $\Omega_{m}^{(0)}$, the quintom model provides a range of possibilities of $H_{0}$, starting from the value corresponding to the $\Lambda$CDM model to higher values. This unique property of the coupled quintom model interestingly makes it a good candidate to resolve the Hubble tension problem. For $H_{0}\simeq 74$ km s$^{-1}$ Mpc$^{-1}$, it corresponds to the range $\delta= 0.11-0.12$.  

%======================================%
%<<<<<<<<<<<< SECTION V  >>>>>>>>>>>>>>%
%======================================%
%%%%%%%%%%%%%%%%%%%%%%%%%%%%%%%%%%%%%%%%%%%%%%%%%%%
%%%%%%%%%%%%%%%%%%%%%%%%%%%%%%%%%%%%%%%%%%%%%%%%%%%
%%%%%%%%%%%%%%%%%%%%%%%%%%%%%%%%%%%%%%%%%%%%%%%%%%%
\section{Comparison with Data and Observational Constraints from the early and late times} 
\label{luminosity}
%%%%%%%%%%%%%%%%%%%%%%%%%%%%%%%%%%%%%%%%%%%%%%%%%%%
%%%%%%%%%%%%%%%%%%%%%%%%%%%%%%%%%%%%%%%%%%%%%%%%%%%

In this section we compare cosmology obtained by the quintom model with observational data, particularly the CMB constraints from the early time $z\geq z_{\text{dec}}$, BAO constraints originated from $z=z_{\text{drag}}$ and Type Ia Supernovae~(SN) from the late time $z\lesssim 2.3$. The Planck constraints are determined from the base $\Lambda$CDM model, therefore certain constraints are not necessarily valid for other models. Since the Hubble tension arises due to the more accurately measured luminosity of SN Ia resulting in larger value of $H_{0}$ than the value implied from the Planck CMB measurement based on the assumption of $\Lambda$CDM model, some of the constraints that depends on the cosmological model could be relaxed, e.g. the constraint $\Omega_{m}^{(0)}h^{2}=0.14170\pm 0.00097$~\cite{Ade:2015xua}. The benchmark quintom model we are considering is based on the choice of parameters that would suppress the difference in the iSW effects in the late time from the $\Lambda$CDM, by tuning the initial conditions and model parameters so that $\Omega_{DE}^{(0)}, \Omega_{m}^{(0)}$ are as close to the best-fit values of $\Lambda$CDM model as possible. The cosmological parameters are given in Table~\ref{presentvalues} obtained from the initial conditions: $x_1 = 1 \times 10^{-5}$, $x_2 = 1 \times 10^{-10}$, $x_3 = 1 \times 10^{-5}$, and $x_4 = 0.9983$ at $N = -13.79$ and model parameters $\lambda_{\phi}=0.10, \delta =0.113$. Another support for this benchmark is the excellent fit with the SN Ia data. Subsequent analysis reveals that $\Omega_{m}^{(0)}\simeq 0.308-0.315$ is prefered for the quintom model with $H_{0}\simeq 73.4$~km s$^{-1}$ Mpc$^{-1}$.  This however gives $\Omega_{m}^{(0)}h^{2}=0.166-0.170$. Here and henceforth, we refer to this quintom model as Quintom I. 

Starting with the SN Ia observations, we use observational data between the magnitude $m_{B}$ and redshift parameter of Type Ia Supernovae from Ref.~\cite{Scolnic:2017caz}~(Pantheon analysis) and take the absolute magnitude $M$ to be a fitting parameter unique for the entire set of data. To focus only on the essential differences between the quintom and $\Lambda$CDM models, the statistical analysis is simplified to contain only one parameter, $M$, which is assumed to include not only the absolute magnitude but also the combined effects of other nuisance parameters such as stretch and color measure of the SN  Ia data~(for more careful analyses including stretch and color measure, see e.g. Ref.\cite{Scolnic:2017caz,Jones:2017udy,Conley:2011ku}). The distance modulus $\mu_{L}$ is related to the observable $m_{B}$ and the luminosity distance $d_{L}$ by
\bea
\mu_{L}&=&m_{B}-M=5\log_{10}\left( \frac{d_{L}}{\text{Mpc}}\right)+25.
\ena
The luminosity distance contains information of the evolution of the Universe through the Hubble parameter,
\bea
d_{L}&=&c(1+z)\int_{0}^{z}~\frac{dz}{H(z)},
\ena
where $H(z)$ can be calculated from Eq.~(\ref{pheneq}) and Eq.~(\ref{HNeq}) depending on the model.  For the $\Lambda$CDM and other non-coupled phenomenological models, Eq.~(\ref{pheneq}) will suffice.  On the other hand, our quintom model with phantom-matter coupling can be more accurately calculated using Eq.~(\ref{HNeq}).
\begin{figure}[h]
	\includegraphics[width=9cm]{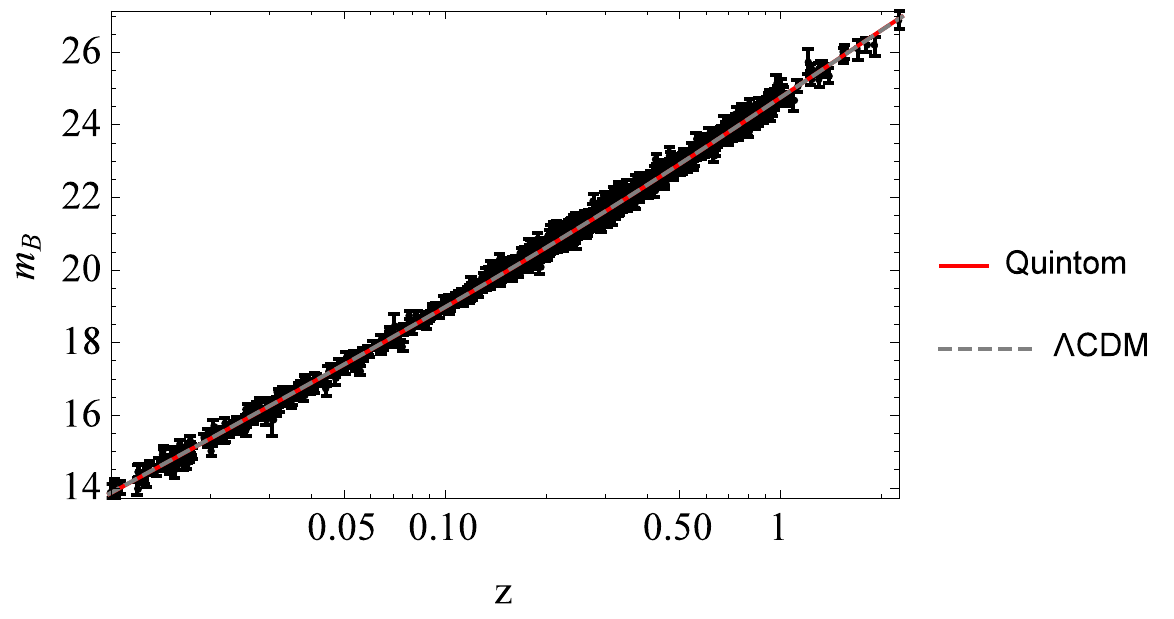} \\
	\includegraphics[width=9cm]{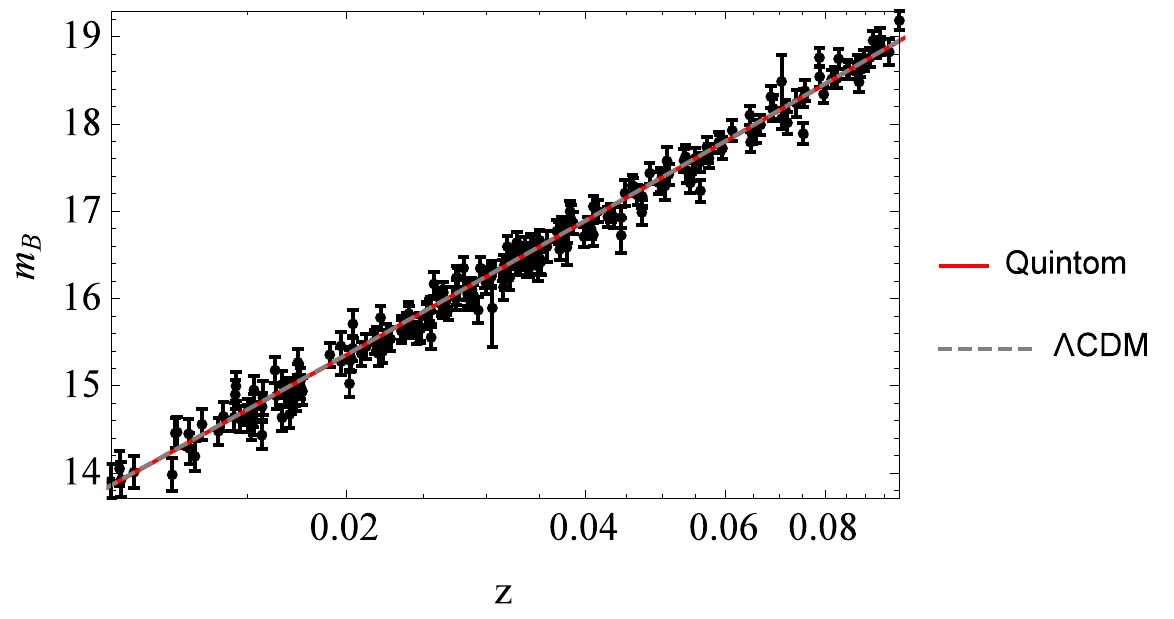}
	\caption{Comparison between Type Ia Supernovae data from Ref.~\cite{Scolnic:2017caz} and theoretical models: $\Lambda$CDM and quintom.}
\label{luminositypic}
\end{figure}
Figure \ref{luminositypic} shows the comparison between theoretical models and the Supernovae observational data. The $\Lambda$CDM parameters are chosen to be $h=67.4, \Omega_{m}^{(0)}=0.308$. The fits of all 1048 data points for $z<2.3$ and of 211 data points for small redshift $z<0.1$ are presented in Fig.~\ref{luminositypic} where the two models appear to be degenerate on a single line.  We define
\be
\chi^{2}\equiv \Delta \vec{m}^{T}C^{-1}\Delta \vec{m},
\ee
where $\Delta \vec{m}_{i}=m_{B,i}^{\rm obs}-m_{B,i}^{\rm th}~(i=1$ to number of data points), $C=D_{\rm stat}+C_{\rm sys}$. The uncertainty matrix $C$ contains diagonal statistical matrix $D_{\rm stat}$ and the off-diagonal covariance matrix $C_{\rm sys}$, see Ref.~\cite{Scolnic:2017caz}.  For all 1048 data points, the chi-square values of the fitting are $\chi^{2}_{\Lambda \rm{CDM}}=1027.07$ for $M=-19.429$ and $\chi^{2}_{\rm{quin}}=1026.74$ for $M=-19.247$.  For 211 low-redshift data points, $\chi^{2}_{\Lambda \rm{CDM}}=217.804$ for $M=-19.442$ and $\chi^{2}_{\rm{quin}}=217.784$ for $M=-19.257$ respectively\footnote{For $\Lambda$CDM with $h=67.4, \Omega_{m}^{(0)}=0.315$~\cite{Aghanim:2018eyx}, the fits have $\chi^{2}_{\Lambda \rm{CDM}}=1027.45~(217.858), M=-19.426~(-19.441)$ for 1048~(211) data points respectively.}. Quintom model clearly provides an equally good fit to the $\Lambda$CDM. 

The degenerate plots of both models are the result of the same values of matter and dark energy densities at late time between the two models~(since we tune the benchmark quintom model so that this is the case), while the difference in $H_{0}$ is compensated by different fitting values of the absolute magnitude $M$.  Quintom model prefers the absolute magnitude around $-19.25$ to $-19.26$, while the $\Lambda$CDM prefers $|M|\simeq 19.43-19.44$. The quintom model gives value of the best-fit $M$ very close to the central value $\overline{M}=-19.25\pm 0.20$ given in Ref.~\cite{Richardson:2014gqa}. However, the error bar is sufficiently large to accommodate the best-fit $M$ of $\Lambda$CDM. More precise measurement of the absolute magnitude of the SN Ia could potentially distinguish which model is more favourable.  

Next we consider the acoustic peaks of the CMB in the quintom model. Generically, the multipole $\ell_{A}$ of the acoustic peaks in the CMB is given by
\be
\ell_{A}=\frac{\pi}{\theta_{A}}=\pi\frac{(1+z_{\rm dec})d_{A}(z_{\rm dec})}{r_{s}(z_{\rm dec})},
\ee
where $\theta_{A}$ is the acoustic angular scale, $d_{A}(z)$ is the angular diameter distance, $r_{s}(z)$ is the comoving sound horizon and $z_{\rm dec}$ is the redshift parameter at the matter-radiation decoupling. $d_{A}(z),r_{s}(z)$ can be calculated from 
\bea
d_{A}(z)&=&\frac{c}{1+z}\int_{0}^{z}\frac{dz}{H(z)}, \\
r_{s}(z) &=& \frac{c}{\sqrt{3}}\int_{z}^{\infty}\frac{dz}{H(z)\sqrt{1+R_{s}(z)}},  \label{rseqn}
\ena
where $R_{s}=\displaystyle{\frac{3\rho_{b}}{4\rho_{\gamma}}}=\displaystyle{\frac{3\Omega_{b}^{(0)}}{4\Omega_{\gamma}^{(0)}(1+z)}}$. From lower figure of Fig.~\ref{evolution}, the phantom contribution $\Omega_{\sigma}$ is smaller than 0.01 or 1 percent throughout the history of the universe and consequently its effect appears only in $H(z)$ at the leading order. The phantom-matter coupling only reduces the matter density very slowly without interfering with the physics of matter-radiation during the transition epoch. Therefore during the radiation-matter transition era, $z_{\rm dec}$ can be approximated using Hu and Sugiyama formula~\cite{Hu:1995en}
\be
z_{\rm dec}=1048~(1 + 0.00124w_{b}^{-0.738})(1 + g_{1}w_{m}^{g_{2}}),
\ee
where $w_{m,b}=\Omega_{m,b}^{(0)}h^{2}$, 
\beann
g_{1} &=& 0.0783w_{b}^{-0.238}/(1 + 39.5w_{b}^{0.763}), \\
g_{2} &=& 0.560/(1 + 21.1w_{b}^{1.81}).
\enann
In our case, we assume the baryon density to be given by $\Omega_{b}^{(0)}h^2=0.02226$~\cite{Ade:2015xua} and photon density by 
\be
\Omega_{\gamma}^{(0)}=\Omega_{r}^{(0)}/(1+0.2271 N_{\rm eff}),
\ee
where $N_{\rm eff}=3.13$ is the effective number of relativistic neutrinos. This gives $z_{\rm dec}=1093.98$. We then numerically calculate the multipole to be $\ell_{A}=285.54$, in a tension with the CMB result from Planck Collaboration that prefers $\ell_{A} \approx 300$.

\begin{figure}[h]
	\includegraphics[width=7.5cm]{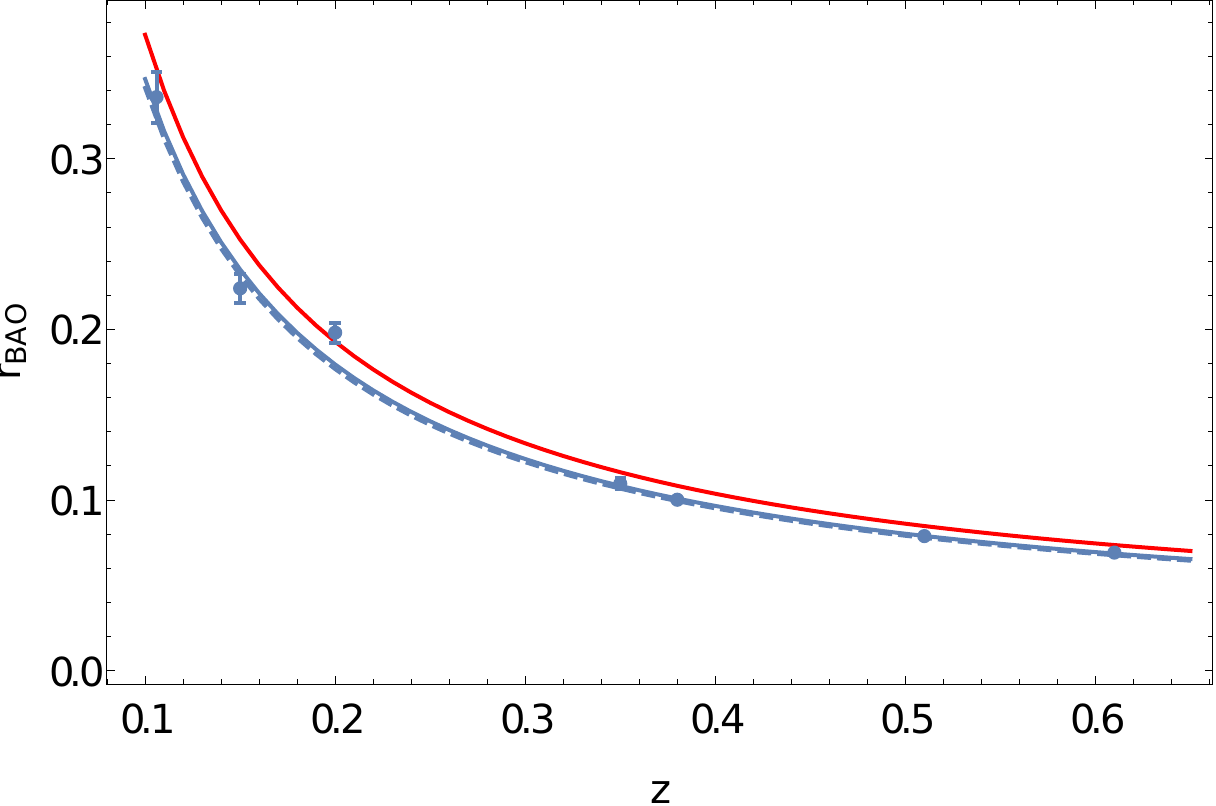}
	\caption{$r_{\rm BAO}$ versus redshift plots in comparison with observational data.  Plot of quintom~($\Lambda$CDM) model is in solid~(dashed), Quintom I~(II) is depicted in red~(blue).}
\label{rBAO}
\end{figure}

Another check of the model is the baryon acoustic oscillations.  The relative BAO distance $r_{\rm BAO}$ can be calculated from
\be
r_{\rm BAO}(z)=\frac{r_{s}(z_{\rm drag})}{[(1+z)^{2}d_{A}^{2}(z)cz/H(z)]^{1/3}}. \label{defrBAO}
\ee
Again, since the fraction of phantom is less than 1 percent and its effect is only to reduce the matter density very slowly, the value of $z_{\rm drag}$ can thus be approximated by the usual Eisenstein and Hu formula~\cite{Eisenstein:1997ik}
\be
z_{\rm drag}= \frac{1291w_{m}^{0.251}}{1 + 0.659w_{m}^{0.828}}(1 + b_{1}w_{b}^{b_{2}}),
\ee
where
\beann
b1 &=& 0.313w_{m}^{-0.419}(1 + 0.607w_{m}^{0.674}), \\
b2 &=& 0.238w_{m}^{0.223}.
\enann
In our quintom model, we adjust the value of $r_{s}(z_{\rm drag})$ by a factor of $1.0275$ to compensate for the discrepancy between the Eisenstein\&Hu formula and the numerical result~\cite{Ade:2013zuv}. With $z_{\rm drag}=1065.71$, the plot of $r_{\rm BAO}$ is shown in Fig.~\ref{rBAO}, observational data are obtained from Ref.~\cite{Percival:2007yw,Beutler et al.(2011),Ross:2014qpa,Alam:2016hwk}. To be consistent with the quintom evolution, the cutoff in the integration limit in Eq.~(\ref{rseqn}) is set to $z_{cut}=\exp(N_{0}-N_{i})-1$ where $N_{i}=-13.79$ for the quintom models.
Both the quintom model and the $\Lambda$CDM fit the BAO observation well~($\Lambda$CDM is better except for $z=0.20$ point).  

\begin{figure}[h]
	\includegraphics[width=9.0cm]{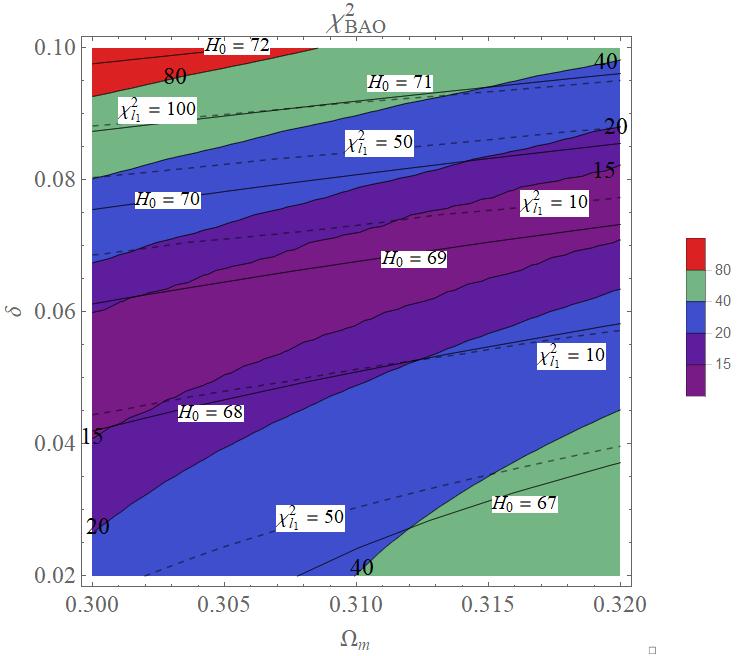}
	\caption{$\chi^{2}$ contours of the quintom model~($\lambda_{\phi}=0.10$). Dashed lines represent $\chi^{2}_{l_{1}}$ of the first acoustic peak $l_{1}$ of the CMB spectrum, non-shaded solid lines represent $H_{0}$ of the quintom model.}
\label{chisqplot}
\end{figure}

In summary, the benchmark model Quintom I provides equally good fit for the SN Ia data, decent fit for the $r_{\rm BAO}$, but slightly small value of $\ell_{A}=285.54$.  The coupled quintom model resolves the Hubble tension but at the same time is in tension with the acoustic peak and BAO measurements. However, the complete scan of the parameter space $(\delta,\Omega_{m}^{(0)})$ of the coupled quintom model shown in Fig.~\ref{chisqplot} reveals that there are regions that yield more satisfactory fits to the BAO data and CMB's first acoustic peak while significantly relieve the tension in $H_{0}$ even though do not completely resolve it. Using the minimization of chi-square of the BAO fitting as the anchor, the benchmark quintom model, Quintom II, is defined with $ \lambda_{\phi}=0.10, \delta=0.06, \Omega_{m}^{(0)}=0.308$. Quintom II fits the SN Ia data with $\chi^{2}_{\rm{quin}}=1026.81$ for $M=-19.3925$ for all 1048 data points. All chi-square values of Quintom II are actually smaller than the $\Lambda$CDM as will be shown below.

\begin{figure}[h]
	\includegraphics[width=8.6cm]{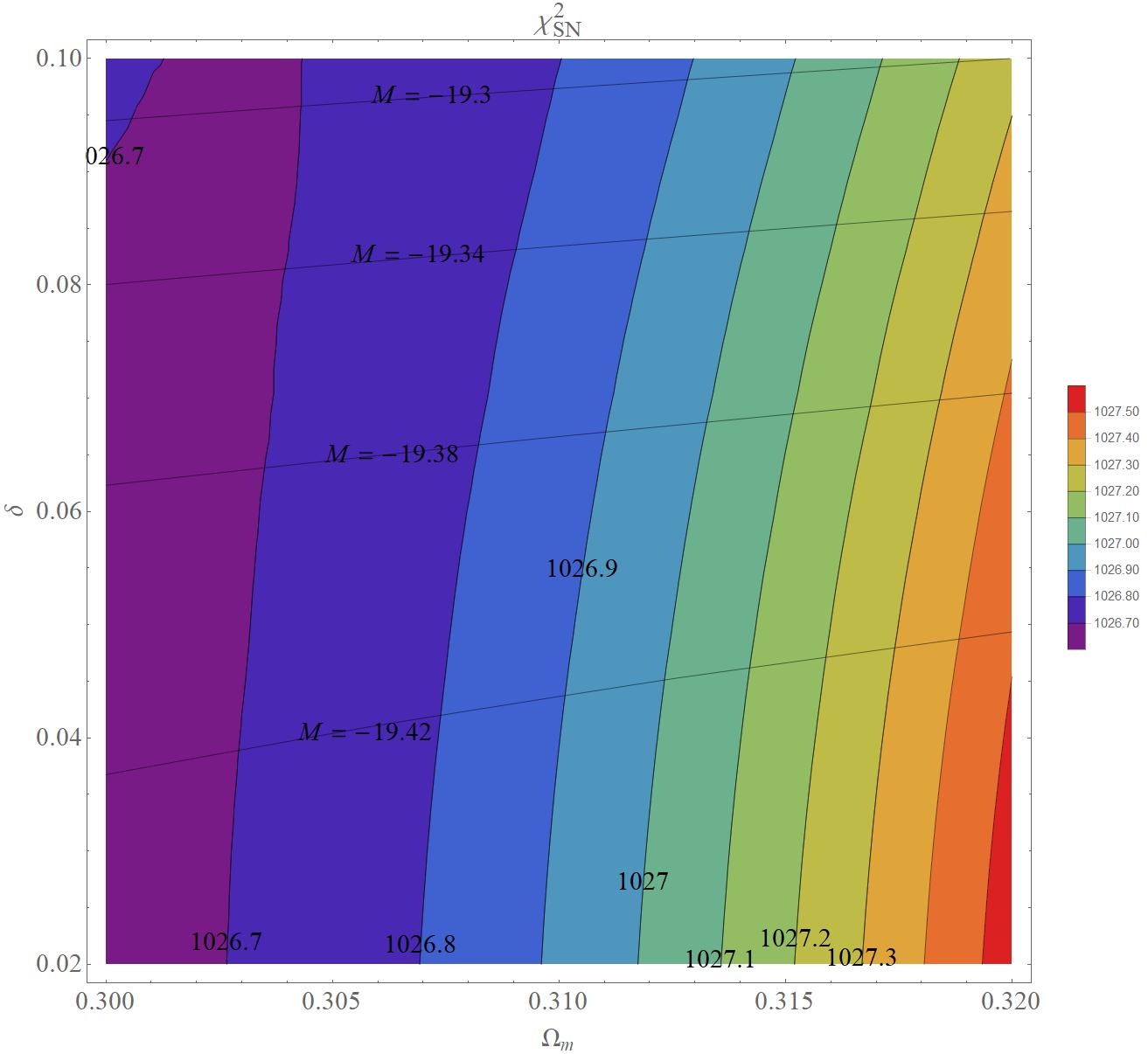}
	\caption{$\chi^{2}$ contours of the quintom model~($\lambda_{\phi}=0.10$) fitting to the SNIa data. Best-fit $M$ contours are depicted as horizontal lines.}
\label{chisqSNplot}
\end{figure}

In addition to Quintom II, Fig.~\ref{chisqplot} shows that the models in the middle region of the parameter space, $\delta=0.02-0.08$, can fit well both with the BAO data and the first CMB peak while giving the present-time Hubble parameter in the range $H_{0}=68-69.5 ~\text{km s}^{-1}\text{Mpc}^{-1}$. Fig.~\ref{chisqSNplot} shows the chi-square contours of the SN Ia fit of the quintom model, the model prefers $\Omega_{m}^{(0)}\simeq 0.30-0.31$ for $\delta =0-0.10$.

In order to find the best model we have to consider a total chi-square of each model. The total chi-square is defined as
\bea
\chi^2 = \chi^2_{\rm SN} + \chi^2_{\rm BAO} + \chi^2_{l_1} \,,
\ena 
where the chi-square of the BAO is given by
\bea
\chi^2_{\rm BAO} = \sum_{i = 1}^{N_{\rm BAO}} \frac{(r_{\rm BAO\,i,th} - r_{\rm BAO\,i,obs})^2}{\sigma_{r_{\rm BAO\,i}}^2} \,.
\ena
The $r_{\rm BAO\,i,th}$ is defined as Eq. (\ref{defrBAO}), while $r_{\rm BAO\,i,obs}$ and $\sigma_{r_{\rm BAO\,i}}$ are observational data and error bars as shown in Fig. \ref{rBAO}. 
%We find 
%\beann
%\chi^2_{\rm BAO} (\rm quintom) &=& 165.371 \,, \\
%\chi^2_{\rm BAO} (\Lambda \rm CDM) &=& 21.2908 \,.
%\enann
For the CMB we consider only the first acoustic peak of the CMB anisotropy, then the chi-square of the first peak is \cite{Arevalo:2016epc}
\bea
\chi^2_{l_1} = \left(\frac{l_{1}^{\rm th} - l_{1}^{ \rm obs}}{\sigma_{l}}\right)^2 \,.
\ena
The $l_{1}$ is the position of the first peak given by~\cite{Hu:2000ti,Doran:2001yw}
\bea
l_1 = l_A (1 - \delta_1) \,,
\ena
where $\delta_1 = 0.267 (r/0.3)^{0.1}$ and $r = \rho_{r} / \rho_{m}$ at $z_{\rm dec}$. From Ref.~\cite{Aghanim:2015xee} the first peak of the TT power spectrum is $l_{1}^{\rm obs} = 220.0$ where $\sigma_{l} = 0.5$. We thus find
\bea
l_1 (\rm Quintom~I) &=& 210.093 \,, \notag\\
l_1 (\rm Quintom~II) &=& 220.357 \,, \notag\\
l_1 (\Lambda \rm CDM) &=& 221.757 \,. \notag
\ena
%which correspond to
%\bea
%\chi^2_{\rm CMB} (\rm quintom) &=& 501.004 \,, \\
%\chi^2_{\rm CMB} (\Lambda \rm CDM) &=& 12.3425 \,.
%\ena
%Therefore, the total chi-square is 
%\bea
%\chi^2 (\rm quintom) &=&  \,, \\
%\chi^2 (\Lambda \rm CDM) &=& 12.3425 \,.
%\ena
The values of chi-square of both models are presented in Table \ref{chisq}.

\begin{table}[ht]
\begin{center}
  \begin{tabular}{|c|c|c|c|c|c|}
\hline
& && & & \\[-.5em]
$ $&$H_{0}$&$\chi^2_{\rm SN}$&$\chi^2_{\rm BAO}$&$\chi^2_{l_1}$&$\chi^2$
\\
& $(\text{km s}^{-1}\text{Mpc}^{-1})$&& & & \\ \hline
& && & & \\[-.5em]
Quintom I&$73.356$&$1026.74$&$161.83$&$392.626$&$1581.20$
\\

($\delta=0.113$)& && & & \\ \hline
& && & & \\[-.5em]
Quintom II&$68.55$&$1026.81$&$13.711$&$0.5104$&$1041.03$
\\[.5em]

($\delta=0.06$)& && & & \\ \hline
& && & & \\[-.5em]
$\Lambda$CDM&$67.4$&$1027.07$&$21.2862$&$12.3425$&$1060.70$
\\[.5em]
\hline
 \end{tabular}
    \caption{Chi-square of the quintom models~($\lambda_{\phi}=0.10$) and the $\Lambda$CDM model.}
\label{chisq}
\end{center}
\end{table}

Since the quintom model has 2 more parameters ($\lambda_{\phi}$ and $\delta$) than the $\Lambda$CDM model, according to the Akaike Information Criterion (AIC) and the Bayesian Information Criterion (BIC) we have to take into account the model parameters and number of data points as \cite{Arevalo:2016epc}
\bea
{\rm AIC} &=& \chi^2_{min} + 2d \,, \\
{\rm BIC} &=& \chi^2_{min} + d \ln N \,,
\ena
where $d$ is the number of the free parameters in the model, $N$ is the number of the data points, and $\chi^2_{min}$ is the minimum value of the chi-square total. The preferred model is the model which has small value of the AIC and BIC. The $\Lambda$CDM model has 4 free parameters~($\Omega_{m}^{(0)}, \Omega_{b}^{(0)}, H_{0},M$) and the data points using in this work is $1048 {\rm (SN)} + 7 {\rm (BAO)} + 1 {\rm (CMB)} = 1056$, then we find
\beann
{\rm AIC ~(Quintom I, II)} &=& 1593.20, 1053.03 \,, \\
{\rm AIC ~(\Lambda CDM)} &=& 1068.70 \,,  ~\\
{\rm BIC ~(Quintom I, II)} &=& 1622.97,~1082.80 \,, \\
{\rm BIC ~(\Lambda CDM)} &=& 1088.55 \,.
\enann
Comparison with three observations (the Type Ia supernovae, the baryons acoustic oscillation, and the first acoustic peak of the CMB anisotropy) indicates that the $\Lambda$CDM model is more~(less) preferred model than Quintom I~(II) respectively. Quintom II relieves Hubble tension and even provide better fits to BAO and CMB $l_{1}$ data.

%======================================%
%<<<<<<<<<<<< SECTION VI  >>>>>>>>>>>>>>%
%======================================%
%%%%%%%%%%%%%%%%%%%%%%%%%%%%%%%%%%%%%%%%%%%%%%%%%%%
%%%%%%%%%%%%%%%%%%%%%%%%%%%%%%%%%%%%%%%%%%%%%%%%%%%
%%%%%%%%%%%%%%%%%%%%%%%%%%%%%%%%%%%%%%%%%%%%%%%%%%%
\section{Conclusions and Discussions} 
\label{conclusions}
%%%%%%%%%%%%%%%%%%%%%%%%%%%%%%%%%%%%%%%%%%%%%%%%%%%
%%%%%%%%%%%%%%%%%%%%%%%%%%%%%%%%%%%%%%%%%%%%%%%%%%%
In this work, the Hubble tension is alleviated by addition of a very small negative density component to the Universe. Such small contribution does not change the values $\Omega_{m}^{(0)}\simeq 0.31, \Omega_{DE}^{(0)}\simeq 0.69$ constrained by the Planck's CMB observation from the early Universe.  As a realization of the idea, we consider a quintom model with conformal phantom-matter coupling and self-interacting quintessence that gives a viable cosmological scenario with the correct density parameters.  The model satisfies the general phenomenological  conditions, i.e., starting with radiation dominated era, continuing with matter and dark energy dominated era subsequently.  It also contains the phantom divide crossing and effectively alleviates the Hubble tension, giving $H_{0}=73.356~{\rm km ~s}^{-1}{\rm Mpc}^{-1}$ and $\Omega_{m}^{(0)}=0.308, \Omega_{\phi}^{(0)}=0.692, \Omega_{\sigma}^{(0)}=-0.00164$ as shown in Table \ref{presentvalues}.

Phenomenologically, as discussed in Section \ref{GenPhen}, the required negative density of the extra component $X$ for $w_{X}=1$ is $\Omega_{X}^{(0)}=-5.247 \times 10^{-11}$, this is based on the non-coupled assumption of $X$ to normal matter.  In our quintom model, the conformal phantom-matter coupling is introduced in order to control the size of the negative density of the phantom field $\sigma$.  In this coupled model, the negative density of the phantom field becomes $\Omega_{\sigma}^{(0)}=-0.00164$ for $w_{\sigma}=1$. 

For the benchmark quintom model that mimic late-time densities of the $\Lambda$CDM, the small redshift~($z<10$) iSW effect originated from the dark energy should be closely similar to the $\Lambda$CDM. We found that the SN Ia fits of the benchmark quintom are as equally good as the fiducial $\Lambda$CDM but with different best-fit absolute magnitude $M$. More precise determination of absolute magnitude of SN Ia in the future observation could potentially distinguish which model is more prefered. The BAO distance fit with observation is decent but the $\Lambda$CDM fit is better except for one point~($z=0.20$).  However, the acoustic peak multipole $\ell_{A}=285.54$ is about 5\% smaller than observation. The benchmark model Quintom I completely resolves tension in the Hubble parameter but is in obvious tension with the peak position of the CMB and the BAO measurements. However, parameter scan of the quintom model reveals region $0.02<\delta<0.10, \Omega_{m}^{(0)}<0.31$ of the parameter space which provides good to excellent fits to the BAO, first acoustic peak of CMB anisotropy and SN Ia data. An example Quintom II model is presented and demonstrated using AIC and BIC that it is a better-fit model than the $\Lambda$CDM and yet significantly alleviates the Hubble tension.

%%%%%%%%%%%%%%%%%%%%%%%%%%%%%%%%%%%%%%%%%%%%%%%%%%%
%%%%%%%%%%%%%%%%%%%%%%%%%%%%%%%%%%%%%%%%%%%%%%%%%%%
\section*{Acknowledgements}
%%%%%%%%%%%%%%%%%%%%%%%%%%%%%%%%%%%%%%%%%%%%%%%%%%%
%%%%%%%%%%%%%%%%%%%%%%%%%%%%%%%%%%%%%%%%%%%%%%%%%%%

We appreciate very helpful suggestions from D.M. Scolnic on the data of SN Ia. S.P. (first author) is supported by Rachadapisek Sompote Fund for Postdoctoral Fellowship, Chulalongkorn University. P.B. is supported in part by the Thailand Research Fund (TRF),
Office of Higher Education Commission (OHEC) and Chulalongkorn University under grant RSA6180002. L.T. is supported by Postdoctoral Fellowship of King Mongkut's University of Technology Thonburi.

%===============================%
%<<<<<<<<<<<< APPENDIX >>>>>>>>>>>>>>%
%======================================%
%%%%%%%%%%%%%%%%%%%%%%%%%%%%%%%%%%%%%%%%%%%
%%%%%%%%%%%%%%%%%%%%%%%%%%%%%%%%%%%%%%%%%%%
%%%%%%%%%%%%%%%%%%%%%%%%%%%%%%%%%%%%%%%%%%%
%%%%%%%%%%%%%%%%%%%%%%%%%%%%%%%%%%%%%%%%%%%
\appendix
%%%%%%%%%%%%%%%%%%%%%%%%%%%%%%%%%%%%%%%%%%%%%%%%%%%
%%%%%%%%%%%%%%%%%%%%%%%%%%%%%%%%%%%%%%%%%%%%%%%%%%%
%%%%%%%%%%%%%%%%%%%%%%%%%%%%%%%%%%%%%%%%%%%%%%%%%%%
\setcounter{table}{0}
\renewcommand{\thetable}{A\arabic{table}}

\section{Fixed points}
\label{fixedpoint}

\begin{widetext}

\begin{table}[thb]
\begin{center}
  \begin{tabular}{|c||c|c|c|c|c|}
\hline 
& & & & & \\[-.5em]
&$x_1$&$x_2$&$x_3$&$x_4$& Existence
\\[.5em]
\hline
& & & & & \\[-.5em]
(a)&$x_1^2 - x_3^2 = 1$&$0$& &$0$&$x_1^2 - x_3^2 = 1$
\\[.5em]
\hline
& & & & & \\[-.5em]
(b)&$0$&$0$&$0$&$1$&All
\\[.5em]
\hline
& & & & & \\[-.5em]
(c)&$0$&$0$ &$- \frac{1}{\sqrt{6}\delta}$&$\sqrt{1+ \frac{1}{2\delta^2}}$&All
\\[.5em]
\hline
& & & & & \\[-.5em]
(d)&$0$&$0$&$\sqrt{\frac{2}{3}}\delta$&$0$&All
\\[.5em]
\hline
& & & & & \\[-.5em]
(e)&$\frac{2\sqrt{6}}{3\lambda_{\phi}}$&$\frac{2\sqrt{3}}{3\lambda_{\phi}}$&$0$&$\sqrt{1 - \frac{4}{\lambda_{\phi}^2}}$&$\lambda_{\phi} \geq 2$
\\[.5em]
\hline
& & & & & \\[-.5em]
(f)&$\frac{2\sqrt{6}}{3\lambda_{\phi}}$&$\frac{2\sqrt{3}}{3\lambda_{\phi}}$&$-\frac{1}{\sqrt{6}\delta}$&$\frac{\sqrt{\lambda_{\phi}^2 + 2\delta^2 (\lambda_{\phi}^2 - 4)}}{\sqrt{2}\delta \lambda_{\phi}}$&$ 0 < \lambda_{\phi} < 2\,, 0 < \delta \leq \sqrt{\frac{\lambda_{\phi}^2}{8 - 2 \lambda_{\phi}^2}}$ \\ 
& & & & &or $ \lambda_{\phi} \geq 2\,, \delta > 0$
\\[.5em]
\hline
& & & & & \\[-.5em]
(g)&$\frac{\lambda_{\phi}}{\sqrt{6}}$&$\sqrt{1 - \frac{\lambda_{\phi}^2}{6}}$&$0$&$0$&$0 < \lambda_{\phi} \leq \sqrt{6}$
\\[.5em]
\hline
& & & & & \\[-.5em]
(h)&$\frac{(3 - 2 \delta^2)\lambda_{\phi}}{\sqrt{6} (\lambda_{\phi}^2 - 2 \delta^2)}$&$\sqrt{\frac{-36\delta^2 + 9\lambda_{\phi}^2 - 4 \delta^4 (\lambda_{\phi}^2 - 6)}{6 (\lambda_{\phi}^2 - 2 \delta^2)^2}}$&$- \sqrt{\frac{2}{3}} \frac{\delta (\lambda_{\phi}^2 - 3)}{2\delta^2 -\lambda_{\phi}^2}$&$0$&$\delta > 0 \,, \lambda_{\phi} = \sqrt{3}$ \\
& & & & &or $\delta > 0 \,, 0 < \lambda_{\phi} \leq \sqrt{3} \,, \sqrt{\frac{6\lambda_{\phi}^2}{6 - \lambda_{\phi}^2}} \geq 2 \delta$ \\
& & & & &or $0 < \delta \leq \sqrt{\frac{3}{2}} \,, \lambda_{\phi} > \sqrt{3}$ \\
& & & & &or $0 < \lambda_{\phi} < \sqrt{3} \,, 2\delta \geq \sqrt{6}$ \\
& & & & &or $\sqrt{3} < \lambda_{\phi} < \sqrt{6} \,, \sqrt{\frac{6\lambda_{\phi}^2}{6 - \lambda_{\phi}^2}} \leq 2\delta$
\\[.5em]
\hline       

 \end{tabular}
    \caption{Fixed points of the autonomous equations (\ref{auto1}) -  (\ref{auto4}).}
\label{fixedpoints}
\end{center}
\end{table}

\end{widetext}

From Table \ref{fixedpoints}, fixed point (a) is a kinetic-dominated point. Radiation dominated epoch can be realized by the fixed point (b), (c), (e), or (f) because $w_{\rm eff} = 1/3$. Fixed point (b) is a standard radiation dominated era, whereas other points are mixture of radiation and other components. Point (d) or (h) can possibly be a matter dominated point, where both of them also have a dark energy component in the matter dominated epoch. The accelerated expansion era can be realized by point (g) or (h). Fixed point (g) is an accelerating expansion fixed point arising in the quintessence model, whereas point (h) is a scaling solution (i.e. a ratio of matter and dark energy is not equal to zero at late-time). Fixed point (d) cannot be an accelerating solution because the dark energy density is not negative at the present.

%%%%%%%%%%%%%%%%%%%%%%%%%%%%%%%%%%%%%%%%%%%%%%%%%%%%%%%%%%%%%%%%%%%%%%%%%%%%%%%%%%%%%%%%%%
\section{Stability analysis}
\label{matrixperturbations}

The autonomous equations can be rewritten as
\beann
\frac{dx_1}{dN} &=& \mathcal{F}(x_1,x_2,x_3,x_4) \,, \nn
\frac{dx_2}{dN} &=&  \mathcal{G}(x_1,x_2,x_3,x_4) \,, \nn
\frac{dx_3}{dN} &=& \mathcal{H}(x_1,x_2,x_3,x_4) \,, \nn
\frac{dx_4}{dN} &=& \mathcal{I}(x_1,x_2,x_3,x_4) \,.
\enann
Stability of the fixed points will be investigated by using the linear perturbation
analysis around each fixed point,
$(x_1^{(c)}, x_2^{(c)}, x_3^{(c)},x_4^{(c)})$, by setting
\beann
x_i(N) &=& x_i^{(c)} + \delta x_i(N) \,, 
\enann
where $i=1,2,3,4$. The perturbation equations then take the form
\bea
\frac{d}{dN} \begin{pmatrix} \delta x_1 \\ \delta x_2 \\ \delta x_3 \\ \delta x_4 \end{pmatrix} =
 \mathcal{M} \begin{pmatrix} \delta x_1 \\ \delta x_2 \\ \delta x_3 \\ \delta x_4 \end{pmatrix} \,, \label{firstordercouple}
\ena
where the matrix $\mathcal{M}$ is given by
\beann
\mathcal{M} = \left.\begin{pmatrix} 
\frac{\partial \mathcal{F}}{\partial x_1} && \frac{\partial \mathcal{F}}{\partial x_2} &&  \frac{\partial \mathcal{F}}{\partial x_3} && \frac{\partial \mathcal{F}}{\partial x_4}\\[.5em]
\frac{\partial \mathcal{G}}{\partial x_1} && \frac{\partial \mathcal{G}}{\partial x_2} &&  \frac{\partial \mathcal{G}}{\partial x_3} && \frac{\partial \mathcal{G}}{\partial x_4}\\[.5em]
\frac{\partial \mathcal{H}}{\partial x_1} && \frac{\partial \mathcal{H}}{\partial x_2} &&  \frac{\partial \mathcal{H}}{\partial x_3} && \frac{\partial \mathcal{H}}{\partial x_4}\\[.5em]
\frac{\partial \mathcal{I}}{\partial x_1} && \frac{\partial \mathcal{I}}{\partial x_2} && \frac{\partial \mathcal{I}}{\partial x_3} && \frac{\partial \mathcal{I}}{\partial x_4}
\end{pmatrix}\right|_{x_1^{(c)},x_2^{(c)},x_3^{(c)},x_4^{(c)}} \,.
\enann

The first order coupled differential equation (\ref{firstordercouple}) has a general solution
\beann
\delta x_i \propto e^{\mu N}\,,
\enann
where $\mu$ is an eigenvalue of the matrix $\mathcal{M}$. Thus, if all eigenvalues are negative (or their real parts are negative for complex eigenvalues), the fixed point is stable.
If at least one eigenvalue is positive, the fixed point is a saddle point. When all of eigenvalues are positive, the fixed point is unstable. 
%Next we will solve the set of autonomous equations by numerical method.

Components of the matrix $\mathcal{M}$ are as follows
\begin{eqnarray*}
\frac{\partial \mathcal{F}}{\partial x_1} &=& \frac{1}{2}(-3 + 9 x_1^2 - 3x_2^2 - 3x_3^2 + x_4^3) \,, \\
\frac{\partial \mathcal{F}}{\partial x_2} &=& -3x_1 x_2 + \sqrt{6} x_2 \lambda_{\phi} \,, 
\frac{\partial \mathcal{F}}{\partial x_3} = -3 x_1 x_3 \,, \\
\frac{\partial \mathcal{F}}{\partial x_4} &=& x_1 x_4 \,, \\
\frac{\partial \mathcal{G}}{\partial x_1} &=& 3x_1 x_2 - \sqrt{\frac{3}{2}} x_2 \lambda_{\phi} \,, \\
\frac{\partial \mathcal{G}}{\partial x_2} &=& \frac{1}{2}(3 + 3x_1^2 - 9 x_2^2 - 3x_3^2 + x_4^2 - \sqrt{6} x_1 \lambda_{\phi}) \,, \\
\frac{\partial \mathcal{G}}{\partial x_3} &=& - 3 x_2 x_3 \,, 
\frac{\partial \mathcal{G}}{\partial x_4} = x_2 x_4 \,, \\
\frac{\partial \mathcal{H}}{\partial x_1} &=& x_1 (3x_3 - \sqrt{6} \delta) \,, 
\frac{\partial \mathcal{H}}{\partial x_2} = - x_2 (3 x_3 + \sqrt{6} \delta) \,, \\
\frac{\partial \mathcal{H}}{\partial x_3} &=& \frac{1}{2}(-3 + 3x_1^2 -3x_2^2 -9x_3^2 + x_4^2 + 2\sqrt{6} x_3 \delta) \,, \nn \\
\frac{\partial \mathcal{H}}{\partial x_4} &=& x_4 (x_3 - \sqrt{6} \delta) \,, \\
\frac{\partial \mathcal{I}}{\partial x_1} &=& 3x_1 x_4 \,, 
\frac{\partial \mathcal{I}}{\partial x_2} = -3 x_2 x_4 \,, 
\frac{\partial \mathcal{I}}{\partial x_3} = -3 x_3 x_4 \,, \\
\frac{\partial \mathcal{I}}{\partial x_4} &=& \frac{1}{2}(-1 + 3x_1^2 -3x_2^2 -3x_3^2 + 3x_4^2) \,.
\end{eqnarray*}

%%%%%%%%%%%%%%%%%%%%%%%%%%%%%%%%%%%%%%%%%%%%%%%%%%
%======================================%
%<<<<<<<<<<<< BIBLIOGRAPHY >>>>>>>>>>>>%
%======================================%
%%%%%%%%%%%%%%%%%%%%%%%%%%%%%%%%%%%%%%%%%%%%%%%%%%
%%%%%%%%%%%%%%%%%%%%%%%%%%%%%%%%%%%%%%%%%%%%%%%%%%


\newpage

\begin{thebibliography}{99}

%\cite{Riess:1998cb}
\bibitem{Riess:1998cb} 
  A.~G.~Riess {\it et al.} [Supernova Search Team],
  %``Observational evidence from supernovae for an accelerating universe and a cosmological constant,''
  Astron.\ J.\  {\bf 116}, 1009 (1998)
  doi:10.1086/300499
  [astro-ph/9805201].
  
%\cite{Perlmutter:1998np}
\bibitem{Perlmutter:1998np} 
  S.~Perlmutter {\it et al.} [Supernova Cosmology Project Collaboration],
  %``Measurements of Omega and Lambda from 42 high redshift supernovae,''
  Astrophys.\ J.\  {\bf 517}, 565 (1999)
  doi:10.1086/307221
  [astro-ph/9812133].
  
%\cite{Horndeski:1974wa}
\bibitem{Horndeski:1974wa} 
  G.~W.~Horndeski,
  %``Second-order scalar-tensor field equations in a four-dimensional space,''
  Int.\ J.\ Theor.\ Phys.\  {\bf 10}, 363 (1974).
  doi:10.1007/BF01807638
  
%\cite{Deffayet:2011gz}
\bibitem{Deffayet:2011gz} 
  C.~Deffayet, X.~Gao, D.~A.~Steer and G.~Zahariade,
  %``From k-essence to generalised Galileons,''
  Phys.\ Rev.\ D {\bf 84}, 064039 (2011)
  doi:10.1103/PhysRevD.84.064039
  [arXiv:1103.3260 [hep-th]].
  
%\cite{Kobayashi:2011nu}
\bibitem{Kobayashi:2011nu} 
  T.~Kobayashi, M.~Yamaguchi and J.~Yokoyama,
  %``Generalized G-inflation: Inflation with the most general second-order field equations,''
  Prog.\ Theor.\ Phys.\  {\bf 126}, 511 (2011)
  doi:10.1143/PTP.126.511
  [arXiv:1105.5723 [hep-th]].
  
%\cite{Heisenberg:2014rta}
\bibitem{Heisenberg:2014rta} 
  L.~Heisenberg,
  %``Generalization of the Proca Action,''
  JCAP {\bf 1405}, 015 (2014)
  doi:10.1088/1475-7516/2014/05/015
  [arXiv:1402.7026 [hep-th]].
  
%\cite{DeFelice:2016yws}
\bibitem{DeFelice:2016yws} 
  A.~De Felice, L.~Heisenberg, R.~Kase, S.~Mukohyama, S.~Tsujikawa and Y.~l.~Zhang,
  %``Cosmology in generalized Proca theories,''
  JCAP {\bf 1606}, no. 06, 048 (2016)
  doi:10.1088/1475-7516/2016/06/048
  [arXiv:1603.05806 [gr-qc]].
  
%\cite{deRham:2010kj}
\bibitem{deRham:2010kj} 
  C.~de Rham, G.~Gabadadze and A.~J.~Tolley,
  %``Resummation of Massive Gravity,''
  Phys.\ Rev.\ Lett.\  {\bf 106}, 231101 (2011)
  doi:10.1103/PhysRevLett.106.231101
  [arXiv:1011.1232 [hep-th]].
  
%\cite{deRham:2010ik}
\bibitem{deRham:2010ik} 
  C.~de Rham and G.~Gabadadze,
  %``Generalization of the Fierz-Pauli Action,''
  Phys.\ Rev.\ D {\bf 82}, 044020 (2010)
  doi:10.1103/PhysRevD.82.044020
  [arXiv:1007.0443 [hep-th]].

%\cite{Abbott:2016blz}
\bibitem{Abbott:2016blz} 
  B.~P.~Abbott {\it et al.} [LIGO Scientific and Virgo Collaborations],
  %``Observation of Gravitational Waves from a Binary Black Hole Merger,''
  Phys.\ Rev.\ Lett.\  {\bf 116}, no. 6, 061102 (2016)
  doi:10.1103/PhysRevLett.116.061102
  [arXiv:1602.03837 [gr-qc]].

%\cite{TheLIGOScientific:2017qsa}
\bibitem{TheLIGOScientific:2017qsa} 
  B.~P.~Abbott {\it et al.} [LIGO Scientific and Virgo Collaborations],
  %``GW170817: Observation of Gravitational Waves from a Binary Neutron Star Inspiral,''
  Phys.\ Rev.\ Lett.\  {\bf 119}, no. 16, 161101 (2017)
  doi:10.1103/PhysRevLett.119.161101
  [arXiv:1710.05832 [gr-qc]].  
    
 %\cite{Baker:2017hug}
\bibitem{Baker:2017hug} 
  T.~Baker, E.~Bellini, P.~G.~Ferreira, M.~Lagos, J.~Noller and I.~Sawicki,
  %``Strong constraints on cosmological gravity from GW170817 and GRB 170817A,''
  Phys.\ Rev.\ Lett.\  {\bf 119}, no. 25, 251301 (2017)
  doi:10.1103/PhysRevLett.119.251301
  [arXiv:1710.06394 [astro-ph.CO]].
  
%\cite{Sakstein:2017xjx}
\bibitem{Sakstein:2017xjx} 
  J.~Sakstein and B.~Jain,
  %``Implications of the Neutron Star Merger GW170817 for Cosmological Scalar-Tensor Theories,''
  Phys.\ Rev.\ Lett.\  {\bf 119}, no. 25, 251303 (2017)
  doi:10.1103/PhysRevLett.119.251303
  [arXiv:1710.05893 [astro-ph.CO]].

%\cite{Ade:2015xua}
\bibitem{Ade:2015xua} 
  P.~A.~R.~Ade {\it et al.} [Planck Collaboration],
  %``Planck 2015 results. XIII. Cosmological parameters,''
  Astron.\ Astrophys.\  {\bf 594}, A13 (2016)
  doi:10.1051/0004-6361/201525830
  [arXiv:1502.01589 [astro-ph.CO]].

%\cite{Riess:2019cxk}
\bibitem{Riess:2019cxk} 
  A.~G.~Riess, S.~Casertano, W.~Yuan, L.~M.~Macri and D.~Scolnic,
  %``Large Magellanic Cloud Cepheid Standards Provide a 1% Foundation for the Determination of the Hubble Constant and Stronger Evidence for Physics beyond $\Lambda$CDM,''
  Astrophys.\ J.\  {\bf 876}, no. 1, 85 (2019)
  doi:10.3847/1538-4357/ab1422
  [arXiv:1903.07603 [astro-ph.CO]].

%\cite{Wong:2019kwg}
\bibitem{Wong:2019kwg} 
  K.~C.~Wong {\it et al.},
  %``H0LiCOW XIII. A 2.4% measurement of $H_{0}$ from lensed quasars: $5.3\sigma$ tension between early and late-Universe probes,''
  arXiv:1907.04869 [astro-ph.CO].

%\cite{Chen:2019ejq}
\bibitem{Chen:2019ejq} 
  G.~C.-F.~Chen {\it et al.},
  %``A SHARP view of H0LiCOW: $H_{0}$ from three time-delay gravitational lens systems with adaptive optics imaging,''
  arXiv:1907.02533 [astro-ph.CO].
  
%\cite{Freedman:2019jwv}
\bibitem{Freedman:2019jwv} 
  W.~L.~Freedman {\it et al.},
  %``The Carnegie-Chicago Hubble Program. VIII. An Independent Determination of the Hubble Constant Based on the Tip of the Red Giant Branch,''
  arXiv:1907.05922 [astro-ph.CO].  

\bibitem{Yuan:2019npk}
W.~Yuan, A.~G.~Riess, L.~M.~Macri, S.~Casertano and D.~Scolnic,
%``Consistent Calibration of the Tip of the Red Giant Branch in the Large Magellanic Cloud on the Hubble Space Telescope Photometric System and a Re-determination of the Hubble Constant,''
Astrophys. J. \textbf{886} (2019), 61
doi:10.3847/1538-4357/ab4bc9
[arXiv:1908.00993 [astro-ph.GA]].

\bibitem{Renk:2017rzu} 
  J.~Renk, M.~Zumalacárregui, F.~Montanari and A.~Barreira,
  %``Galileon gravity in light of iSW, CMB, BAO and H$_0$ data,''
  JCAP {\bf 1710}, no. 10, 020 (2017)
  doi:10.1088/1475-7516/2017/10/020
  [arXiv:1707.02263 [astro-ph.CO]].

\bibitem{Nunes:2018xbm} 
  R.~C.~Nunes,
  %``Structure formation in $f(T)$ gravity and a solution for $H_0$ tension,''
  JCAP {\bf 1805}, no. 05, 052 (2018)
  doi:10.1088/1475-7516/2018/05/052
  [arXiv:1802.02281 [gr-qc]].

\bibitem{Desmond:2019ygn} 
  H.~Desmond, B.~Jain and J.~Sakstein,
  %``A local resolution of the Hubble tension: The impact of screened fifth forces on the cosmic distance ladder,''
  arXiv:1907.03778 [astro-ph.CO].

\bibitem{Alam:2016wpf} 
  U.~Alam, S.~Bag and V.~Sahni,
  %``Constraining the Cosmology of the Phantom Brane using Distance Measures,''
  Phys.\ Rev.\ D {\bf 95}, no. 2, 023524 (2017)
  doi:10.1103/PhysRevD.95.023524
  [arXiv:1605.04707 [astro-ph.CO]].
      
\bibitem{Khosravi:2017hfi} 
  N.~Khosravi, S.~Baghram, N.~Afshordi and N.~Altamirano,
  %``$H_0$ tension as a hint for a transition in gravitational theory,''
  Phys.\ Rev.\ D {\bf 99}, no. 10, 103526 (2019)
  doi:10.1103/PhysRevD.99.103526
  [arXiv:1710.09366 [astro-ph.CO]].

\bibitem{DiValentino:2017rcr} 
  E.~Di Valentino, E.~V.~Linder and A.~Melchiorri,
  %``Vacuum phase transition solves the $H_0$ tension,''
  Phys.\ Rev.\ D {\bf 97}, no. 4, 043528 (2018)
  doi:10.1103/PhysRevD.97.043528
  [arXiv:1710.02153 [astro-ph.CO]].    

\bibitem{Banihashemi:2018has} 
  A.~Banihashemi, N.~Khosravi and A.~H.~Shirazi,
  %``Ginzburg-Landau Theory of Dark Energy: A Framework to Study Both Temporal and Spatial Cosmological Tensions Simultaneously,''
  Phys.\ Rev.\ D {\bf 99}, no. 8, 083509 (2019)
  doi:10.1103/PhysRevD.99.083509
  [arXiv:1810.11007 [astro-ph.CO]].
    
%\cite{Poulin:2018cxd}
\bibitem{Poulin:2018cxd} 
  V.~Poulin, T.~L.~Smith, T.~Karwal and M.~Kamionkowski,
  %``Early Dark Energy Can Resolve The Hubble Tension,''
  Phys.\ Rev.\ Lett.\  {\bf 122}, no. 22, 221301 (2019)
  doi:10.1103/PhysRevLett.122.221301
  [arXiv:1811.04083 [astro-ph.CO]].
  
%\cite{Vattis:2019efj}
\bibitem{Vattis:2019efj} 
  K.~Vattis, S.~M.~Koushiappas and A.~Loeb,
  %``Dark matter decaying in the late Universe can relieve the H0 tension,''
  Phys.\ Rev.\ D {\bf 99}, no. 12, 121302 (2019)
  doi:10.1103/PhysRevD.99.121302
  [arXiv:1903.06220 [astro-ph.CO]].

\bibitem{Kumar:2019wfs} 
  S.~Kumar, R.~C.~Nunes and S.~K.~Yadav,
  %``Dark sector interaction: a remedy of the tensions between CMB and LSS data,''
  Eur.\ Phys.\ J.\ C {\bf 79}, no. 7, 576 (2019)
  doi:10.1140/epjc/s10052-019-7087-7
  [arXiv:1903.04865 [astro-ph.CO]].

\bibitem{Yang:2018euj} 
  W.~Yang, S.~Pan, E.~Di Valentino, R.~C.~Nunes, S.~Vagnozzi and D.~F.~Mota,
  %``Tale of stable interacting dark energy, observational signatures, and the $H_0$ tension,''
  JCAP {\bf 1809}, no. 09, 019 (2018)
  doi:10.1088/1475-7516/2018/09/019
  [arXiv:1805.08252 [astro-ph.CO]].
    
\bibitem{Kreisch:2019yzn} 
  C.~D.~Kreisch, F.~Y.~Cyr-Racine and O.~Doré,
  %``The Neutrino Puzzle: Anomalies, Interactions, and Cosmological Tensions,''
  arXiv:1902.00534 [astro-ph.CO].

\bibitem{Li:2019yem} 
  X.~Li and A.~Shafieloo,
  %``Phenomenologically Emergent Dark Energy and Ruling Out Cosmological Constant,''
  arXiv:1906.08275 [astro-ph.CO].
      
%\cite{Dutta:2018vmq}
\bibitem{Dutta:2018vmq} 
  K.~Dutta, Ruchika, A.~Roy, A.~A.~Sen and M.~M.~Sheikh-Jabbari,
  %``Beyond $\Lambda$CDM with Low and High Redshift Data: Implications for Dark Energy,''
  arXiv:1808.06623 [astro-ph.CO].

\bibitem{Visinelli:2019qqu} 
  L.~Visinelli, S.~Vagnozzi and U.~Danielsson,
  %``Revisiting a negative cosmological constant from low-redshift data,''
  Symmetry {\bf 11}, 1035 (2019)
  doi:10.3390/sym11081035
  [arXiv:1907.07953 [astro-ph.CO]].
  
\bibitem{Panpanich:2018cxo} 
  S.~Panpanich and P.~Burikham,
  %``Fitting rotation curves of galaxies by de Rham-Gabadadze-Tolley massive gravity,''
  Phys.\ Rev.\ D {\bf 98}, no. 6, 064008 (2018)
  doi:10.1103/PhysRevD.98.064008
  [arXiv:1806.06271 [gr-qc]].
    
%\cite{Guo:2004fq}
\bibitem{Guo:2004fq} 
  Z.~K.~Guo, Y.~S.~Piao, X.~M.~Zhang and Y.~Z.~Zhang,
  %``Cosmological evolution of a quintom model of dark energy,''
  Phys.\ Lett.\ B {\bf 608}, 177 (2005)
  doi:10.1016/j.physletb.2005.01.017
  [astro-ph/0410654].

\bibitem{Elizalde:2004mq} 
  E.~Elizalde, S.~Nojiri and S.~D.~Odintsov,
  %``Late-time cosmology in (phantom) scalar-tensor theory: Dark energy and the cosmic speed-up,''
  Phys.\ Rev.\ D {\bf 70}, 043539 (2004)
  doi:10.1103/PhysRevD.70.043539
  [hep-th/0405034].

\bibitem{Elizalde:2008yf} 
  E.~Elizalde, S.~Nojiri, S.~D.~Odintsov, D.~Saez-Gomez and V.~Faraoni,
  %``Reconstructing the universe history, from inflation to acceleration, with phantom and canonical scalar fields,''
  Phys.\ Rev.\ D {\bf 77}, 106005 (2008)
  doi:10.1103/PhysRevD.77.106005
  [arXiv:0803.1311 [hep-th]].

\bibitem{Alexander:2019rsc} 
  S.~Alexander and E.~McDonough,
  %``Axion-Dilaton Destabilization and the Hubble Tension,''
  Physics Letters B 797 (2019)
  doi:10.1016/j.physletb.2019.134830
  [arXiv:1904.08912 [astro-ph.CO]].

\bibitem{Colgain:2019joh} 
  E.~Ó.~Colgáin and H.~Yavartanoo,
  %``Testing the Swampland: $H_0$ tension,''
  arXiv:1905.02555 [astro-ph.CO].
          
%\cite{Naruko:2015zze}
\bibitem{Naruko:2015zze} 
  A.~Naruko, D.~Yoshida and S.~Mukohyama,
  %``Gravitational scalar?tensor theory,''
  Class.\ Quant.\ Grav.\  {\bf 33}, no. 9, 09LT01 (2016)
  doi:10.1088/0264-9381/33/9/09LT01
  [arXiv:1512.06977 [gr-qc]].
 
%\cite{Saridakis:2016ahq}
\bibitem{Saridakis:2016ahq} 
  E.~N.~Saridakis and M.~Tsoukalas,
  %``Cosmology in new gravitational scalar-tensor theories,''
  Phys.\ Rev.\ D {\bf 93}, no. 12, 124032 (2016)
  doi:10.1103/PhysRevD.93.124032
  [arXiv:1601.06734 [gr-qc]].
  
%\cite{Aghanim:2018eyx}
\bibitem{Aghanim:2018eyx} 
  N.~Aghanim {\it et al.} [Planck Collaboration],
  %``Planck 2018 results. VI. Cosmological parameters,''
  arXiv:1807.06209 [astro-ph.CO].
  
%\cite{Amendola:1999er}
\bibitem{Amendola:1999er} 
  L.~Amendola,
  %``Coupled quintessence,''
  Phys.\ Rev.\ D {\bf 62}, 043511 (2000)
  doi:10.1103/PhysRevD.62.043511
  [astro-ph/9908023].
  
%\cite{Tsujikawa:2010sc}
\bibitem{Tsujikawa:2010sc} 
  S.~Tsujikawa,
  %``Dark energy: investigation and modeling,''
  doi:10.1007/978-90-481-8685-3\_8
  arXiv:1004.1493 [astro-ph.CO].
  
%\cite{Hinterbichler:2011ca}
\bibitem{Hinterbichler:2011ca} 
  K.~Hinterbichler, J.~Khoury, A.~Levy and A.~Matas,
  %``Symmetron Cosmology,''
  Phys.\ Rev.\ D {\bf 84}, 103521 (2011)
  doi:10.1103/PhysRevD.84.103521
  [arXiv:1107.2112 [astro-ph.CO]].
  
%\cite{Khoury:2003aq}
\bibitem{Khoury:2003aq} 
  J.~Khoury and A.~Weltman,
  %``Chameleon fields: Awaiting surprises for tests of gravity in space,''
  Phys.\ Rev.\ Lett.\  {\bf 93}, 171104 (2004)
  doi:10.1103/PhysRevLett.93.171104
  [astro-ph/0309300].
   
%\cite{Khoury:2003rn}
\bibitem{Khoury:2003rn} 
  J.~Khoury and A.~Weltman,
  %``Chameleon cosmology,''
  Phys.\ Rev.\ D {\bf 69}, 044026 (2004)
  doi:10.1103/PhysRevD.69.044026
  [astro-ph/0309411].

\bibitem{Yu:2017iju}
H.~Yu, B.~Ratra and F.~Wang,
%``Hubble Parameter and Baryon Acoustic Oscillation Measurement Constraints on the Hubble Constant, the Deviation from the Spatially Flat ΛCDM Model, the Deceleration\UTF{2013}Acceleration Transition Redshift, and Spatial Curvature,''
Astrophys. J. \textbf{856}, no.1, 3 (2018)
doi:10.3847/1538-4357/aab0a2
[arXiv:1711.03437 [astro-ph.CO]].

%\cite{Scolnic:2017caz}
\bibitem{Scolnic:2017caz} 
  D.~M.~Scolnic {\it et al.},
  %``The Complete Light-curve Sample of Spectroscopically Confirmed SNe Ia from Pan-STARRS1 and Cosmological Constraints from the Combined Pantheon Sample,''
  Astrophys.\ J.\  {\bf 859}, no. 2, 101 (2018)
  doi:10.3847/1538-4357/aab9bb
  [arXiv:1710.00845 [astro-ph.CO]].  
  
\bibitem{Jones:2017udy}
D.~Jones, D.~Scolnic, A.~Riess, A.~Rest, R.~Kirshner, E.~Berger, R.~Kessler, Y.~Pan, R.~Foley, R.~Chornock, C.~Ortega, P.~Challis, W.~Burgett, K.~Chambers, P.~Draper, H.~Flewelling, M.~Huber, N.~Kaiser, R.~Kudritzki, N.~Metcalfe, J.~Tonry, R.~Wainscoat, C.~Waters, E.~Gall, R.~Kotak, M.~McCrum, S.~Smartt and K.~Smith,
%``Measuring Dark Energy Properties with Photometrically Classified Pan-STARRS Supernovae. II. Cosmological Parameters,''
Astrophys. J. \textbf{857}, no.1, 51 (2018)
doi:10.3847/1538-4357/aab6b1
[arXiv:1710.00846 [astro-ph.CO]].

\bibitem{Conley:2011ku}
A.~Conley \textit{et al.} [SNLS],
%``Supernova Constraints and Systematic Uncertainties from the First 3 Years of the Supernova Legacy Survey,''
Astrophys. J. Suppl. \textbf{192}, 1 (2011)
doi:10.1088/0067-0049/192/1/1
[arXiv:1104.1443 [astro-ph.CO]].

\bibitem{Richardson:2014gqa} 
  D.~Richardson, R.~L.~Jenkins, J.~Wright and L.~Maddox,
  %``Absolute-Magnitude Distributions of Supernovae,''
  Astron.\ J.\  {\bf 147}, 118 (2014)
  doi:10.1088/0004-6256/147/5/118
  [arXiv:1403.5755 [astro-ph.SR]].
    
\bibitem{Hu:1995en} 
  W.~Hu and N.~Sugiyama,
  %``Small scale cosmological perturbations: An Analytic approach,''
  Astrophys.\ J.\  {\bf 471}, 542 (1996)
  doi:10.1086/177989
  [astro-ph/9510117].  
  
\bibitem{Eisenstein:1997ik} 
  D.~J.~Eisenstein and W.~Hu,
  %``Baryonic features in the matter transfer function,''
  Astrophys.\ J.\  {\bf 496}, 605 (1998)
  doi:10.1086/305424
  [astro-ph/9709112].  
  
\bibitem{Ade:2013zuv} 
  P.~A.~R.~Ade {\it et al.} [Planck Collaboration],
  %``Planck 2013 results. XVI. Cosmological parameters,''
  Astron.\ Astrophys.\  {\bf 571}, A16 (2014)
  doi:10.1051/0004-6361/201321591
  [arXiv:1303.5076 [astro-ph.CO]].  

\bibitem{Percival:2007yw} 
  W.~J.~Percival, S.~Cole, D.~J.~Eisenstein, R.~C.~Nichol, J.~A.~Peacock, A.~C.~Pope and A.~S.~Szalay,
  %``Measuring the Baryon Acoustic Oscillation scale using the SDSS and 2dFGRS,''
  Mon.\ Not.\ Roy.\ Astron.\ Soc.\  {\bf 381}, 1053 (2007)
  doi:10.1111/j.1365-2966.2007.12268.x
  [arXiv:0705.3323 [astro-ph]].
    
\bibitem{Beutler et al.(2011)}
Beutler, F., Blake, C., Colless, M. {\it et al.},
Mon.\ Not.\ Roy.\ Astron.\ Soc.\  {\bf 416}, 3017 (2011).

\bibitem{Ross:2014qpa} 
  A.~J.~Ross, L.~Samushia, C.~Howlett, W.~J.~Percival, A.~Burden and M.~Manera,
  %``The clustering of the SDSS DR7 main Galaxy sample \UTF{2013} I. A 4 per cent distance measure at $z = 0.15$,''
  Mon.\ Not.\ Roy.\ Astron.\ Soc.\  {\bf 449}, no. 1, 835 (2015)
  doi:10.1093/mnras/stv154
  [arXiv:1409.3242 [astro-ph.CO]].

\bibitem{Alam:2016hwk} 
  S.~Alam {\it et al.} [BOSS Collaboration],
  %``The clustering of galaxies in the completed SDSS-III Baryon Oscillation Spectroscopic Survey: cosmological analysis of the DR12 galaxy sample,''
  Mon.\ Not.\ Roy.\ Astron.\ Soc.\  {\bf 470}, no. 3, 2617 (2017)
  doi:10.1093/mnras/stx721
  [arXiv:1607.03155 [astro-ph.CO]].  
  
%\cite{Arevalo:2016epc}
\bibitem{Arevalo:2016epc}
F.~Arevalo, A.~Cid and J.~Moya,
%``AIC and BIC for cosmological interacting scenarios,''
Eur. Phys. J. C \textbf{77}, no.8, 565 (2017)
doi:10.1140/epjc/s10052-017-5128-7
[arXiv:1610.09330 [astro-ph.CO]].
%15 citations counted in INSPIRE as of 25 Aug 2020

\bibitem{Hu:2000ti}
W.~Hu, M.~Fukugita, M.~Zaldarriaga and M.~Tegmark,
%``CMB observables and their cosmological implications,''
Astrophys. J. \textbf{549} (2001), 669
doi:10.1086/319449
[arXiv:astro-ph/0006436 [astro-ph]].

\bibitem{Doran:2001yw}
M.~Doran and M.~Lilley,
%``The Location of CMB peaks in a universe with dark energy,''
Mon. Not. Roy. Astron. Soc. \textbf{330} (2002), 965-970
doi:10.1046/j.1365-8711.2002.05144.x
[arXiv:astro-ph/0104486 [astro-ph]].

%\cite{Aghanim:2015xee}
\bibitem{Aghanim:2015xee}
N.~Aghanim \textit{et al.} [Planck],
%``Planck 2015 results. XI. CMB power spectra, likelihoods, and robustness of parameters,''
Astron. Astrophys. \textbf{594}, A11 (2016)
doi:10.1051/0004-6361/201526926
[arXiv:1507.02704 [astro-ph.CO]].
%752 citations counted in INSPIRE as of 25 Aug 2020

\end{thebibliography}
\end{document}